\documentclass{aa}

\usepackage[]{graphicx}
\usepackage[]{graphics}
\usepackage[]{psfig}
\usepackage[]{epsfig}
\usepackage[]{longtable}
\usepackage[]{times}
\include{epsf}

\begin{document}
\input{psfig.sty}

\def\fdeg{\hbox{$.\mkern-4mu^\circ$}}
\def\farcs{\hbox{$.\!\!^{\prime\prime}$}}
\def\farcm{\hbox{$.\mkern-4mu^\prime$}}
\def\degr{\hbox{$^\circ$}}
\def\arcmin{\hbox{$^\prime$}}
\def\arcsec{\hbox{$^{\prime\prime}$}}
\def\sun{\hbox{$\odot$}}
\def\logg{\hbox{$\log g$}}
\def\Teff{\hbox{$T_{\rm eff}$}}
\def\fh{\hbox{$.\!\!^{\rm h}$}}
\def\fm{\hbox{$.\!\!^{\rm m}$}}
\def\fs{\hbox{$.\!\!^{\rm s}$}}
\def\hou{^{\rm h}}
\def\min{^{\rm m}}
\def\ssec{^{\rm s}}

\def\logg{log\hspace*{1mm}$g$}
\def\mfe{$\langle{\rm Fe}\rangle$}
\def\mh{$\langle{\rm H}\rangle$}


\title{Probable member stars of the gravitational theory-testing globular
clusters AM\,1, Pal\,3 and Pal\,14\thanks{Based on observations obtained at
the European Southern Observatory, Chile (Observing Programme 074.D--0187).}}

\author {Michael Hilker \inst{1}}

\offprints {M.~Hilker}
\mail{mhilker@astro.uni-bonn.de}

\institute{
Sternwarte der Universit\"at Bonn, Auf dem H\"ugel 71, 53121 Bonn, Germany
}

\date{Received 10 October 2005 / Accepted 17 October 2005}

\titlerunning{Probable member stars of AM\,1, Pal\,3 and Pal\,14}

\authorrunning{M.~Hilker}

\abstract{Some of the Galactic outer halo globular clusters are excellent tools
to probe gravitational theories in the regime of weak accelerations (Baumgardt 
et al. \cite{baum05}). The measurement of the line-of-sight velocity dispersion
among stars in these clusters will differentiate between the validity of 
Newtonian dynamics (low velocity dispersion) and the possiblity of modified 
Newtonian dynamics (MOND) or dark matter dominated globular clusters (high 
velocity dispersion). 
In this paper, the properties of probable member stars of the three best-case 
gravitational theory-testing clusters AM\,1, Pal\,3 and Pal\,14 are presented. 
The member selection is based on VLT photometry in Johnson $BV$. The positions
of the stars were determined with an accuracy of the order $\leq0\farcs2$, 
allowing their direct use for follow-up spectroscopy. The distance, reddening, 
age, and metallicities of the clusters were estimated from isochrone fitting.
Furthermore, improved structural parameters, like central coordinates, 
ellipticity, half-light radius, King model core and tidal radius, are 
presented.
}

\maketitle



\section{Introduction}\label{intro}

In the distant halo of our Milky Way there exist several low mass, often
diffuse globular clusters (GCs). Most of these outer halo GCs belong to the 
so-called `young halo clusters', comprising about 30 Milky Way GCs (e.g. 
Zinn \cite{zinn93}, Mackey \& Gilmore \cite{mack04}). From their horizontal 
branch morphology it was estimated that they are about 1-2 Gyr younger than the
`old halo clusters'.  Also they have on average larger half-light radii than 
the inner halo GCs (Mackey \& van~den~Bergh \cite{mack05}).
The similarity of their properties to those of external globular clusters 
(belonging to Milky Way satellite galaxies) led to the idea that they were 
accreted into the Galactic halo (e.g. van~den~Bergh \& Mackey \cite{vdbe04}, 
originally proposed by Searle \& Zinn \cite{sear78}).
The most remote halo GCs (at galacto-centric distances of $\geq100$ kpc) live
on eccentric orbits, mostly undisturbed from external tidal forces of our
Galaxy. 

Recently, it has been proposed by Baumgardt et al. (\cite{baum05}) that some
of these
distant Galactic globular clusters are excellent tools to probe gravitational 
theories in the regime of very weak accelerations. Their internal and external
accelerations both are significantly below the critical acceleration parameter
$a_0$ of modified Newtonian dynamics (MOND, Milgrom \cite{milg83}, Bekenstein 
\& Milgrom \cite{beke84}). In case of MOND the internal velocity dispersion 
among the stars in these clusters would be significantly higher than in the 
Newtonian standard dynamics. A higher than Newtonian velocity dispersion could
also be explained by the existence of dark matter in globular clusters (e.g.
Mashchenko \& Sills \cite{mash05}). The line-of-sight velocity dispersion 
of the 8 gravitational theory-testing GCs listed by Baumgardt et al. 
(\cite{baum05}) is 
expected to range between 0.5 and 1.1\,km$^2$/sec$^2$ in the Newtonian case,
whereas two times higher values (1.1 to 2.4\,km$^2$/sec$^2$) are 
predicted for MOND. Such velocity differences can be measured with existing 
high resolution spectrographs, providing the observation of a statistical 
meaningful sample ($\sim20$-30) of stars.

However, at the distance of the proposed GCs of between 28 and 122 kpc,
high resolution spectroscopic observations of these feeble, low mass clusters 
are challenging. There exist only few evolved stars that are bright enough
to be observed in reasonable integration times with 8-meter class telescopes.
Therefore, it is very useful to know the positions and photometric properties 
of these bright member stars beforehand. 

\begin{table*}[t!]
\caption{\label{tab1} Observation log. The appendices `N' and `S' in the
GC names denote the pointings North and South of their centers,
respectively.}
\begin{tabular}{lccccccccc}
\hline
 & & & & & & & & & \\[-3mm]
Name & $\alpha$(2000)$^a$ & $\delta$(2000)$^a$ & Date & 
\multicolumn{2}{c}{Exposure times [s]} & \multicolumn{2}{c}{Seeing [$\arcsec$]}
 & \multicolumn{2}{c}{Airmass} \\
 & [h:m:s] & [$^\circ$:$\arcmin$:$\arcsec$] & & $B$ & $V$ & $B$ & $V$ & $B$ & 
$V$ \\
\hline
 & & & & & & & & & \\[-3mm]
AM\,1-N & 03:55:02.6 & $-$49:34:41.9 & 2005~Feb~15 & $3\times25$ & $3\times10$
 & 0.6 & 0.7 & 1.20 & 1.20 \\
AM\,1-S & 03:55:02.6 & $-$49:38:42.0 & 2005~Feb~16 & $3\times30$ & $3\times15$
 & 0.6 & 0.6 & 1.21 & 1.22 \\
 & & & & & & & & & \\[-3mm]
Pal\,3-N & 10:05:31.4 & $+$00:06:25.9 & 2005~Feb~15 & $3\times25$ & $3\times10$ 
 & 0.6 & 0.7 & 1.24 & 1.23 \\
Pal\,3-S & 10:05:31.4 & $+$00:02:26.0 & 2005~Feb~15 & $3\times25$ & $3\times10$
 & 0.5 & 0.5 & 1.22 & 1.21 \\
 & & & & & & & & & \\[-3mm]
Pal\,14-N & 16:11:05.0 & $+$14:59:38.0 & 2005~Feb~15 & $3\times25$ & $3\times10$
 & 0.6 & 0.7 & 1.46 & 1.45 \\
Pal\,14-S & 16:11:05.0 & $+$14:55:37.9 & 2005~Feb~15 & $3\times25$ & $3\times10$
 & 0.5 & 0.6 & 1.43 & 1.42 \\
\hline
\end{tabular}
\vskip0.1cm
$^a$The coordinates indicate the pointing of the observations
\end{table*}

For three of the Baumgardt-sample GCs we present those properties
in this paper, namely AM\,1, Pal\,3 and Pal\,14. These clusters were selected
because they are expected to show the largest differences between Newtonian and
MOND dynamics. They are therefore the best cases for testing gravitational
theories. Their basic parameters are given in the Tables~\ref{tab3} and
~\ref{tab4} (most of these parameters were taken from the 2003 version of
Harris' list \cite{harr96}). 
All three clusters belong to the 'young halo' GCs with iron abundances between
$-1.5$ and $-1.8$ dex, and half-light radii in the range 17-25 pc.
An age estimate for Pal\,3 was derived from the detailed study of the colour 
magnitude
diagram, especially the horizontal branch (HB) morphology. It seems that Pal\,3
is between 1 Gyr (VandenBerg \cite{vand00}, Catelan et al. \cite{cate01}) and 2
Gyr (Stetson et al. \cite{stet99}) younger than ordinary inner halo GCs like
M\,3. Also for Pal\,14 a 3 to 4 Gyr younger age than for typical GCs was
estimated from the position of the main sequence turnoff (MSTO) relative 
to the HB level (Sarajedini \cite{sara97}).

\begin{table}[t!]
\caption{\label{tab2} Calibration coefficients for both nights. The zero points
apply for magnitudes transformed from intensities in counts/sec. The signs are 
such that the
following calibration equation applies: $mag_{\rm cal} = mag_{\rm inst} + ZP +
k \times AM + CT \times (B-V)_{\rm cal}$, with ZP = zero point, k = extinction
coefficient, AM = airmass, and CT = colour term.}
\begin{tabular}{l@{\hspace{2mm}}c@{\hspace{2mm}}rrr}
\hline
 & & & & \\[-3mm]
Chip & Filter & zero point & ext. coeff. & colour term \\
\hline
 & & & & \\[-3mm]
chip1 & $V$ & 27.852$\pm$0.002 & $-0.123\pm$0.011 & 0.034$\pm$0.008 \\
      & $B$ & 27.295$\pm$0.009 & $-0.185\pm$0.016 & $-0.031\pm$0.006 \\
 & & & & \\[-3mm]
chip2 & $V$ & 27.841$\pm$0.012 & $-0.162\pm$0.018 & 0.020$\pm$0.004 \\
      & $B$ & 27.298$\pm$0.008 & $-0.208\pm$0.016 & $-0.033\pm$0.006 \\
\hline
\end{tabular}
\end{table}

In this paper, we provide accurate positions, $V$ magnitudes and $(B-V)$ 
colours of the evolved stars in the three clusters AM\,1, Pal\,3 and
Pal\,14. This photometry is necessary and very useful for spectroscopic 
follow-up observations of these best-case gravitational theory-testing
clusters.

\begin{figure}
\psfig{figure=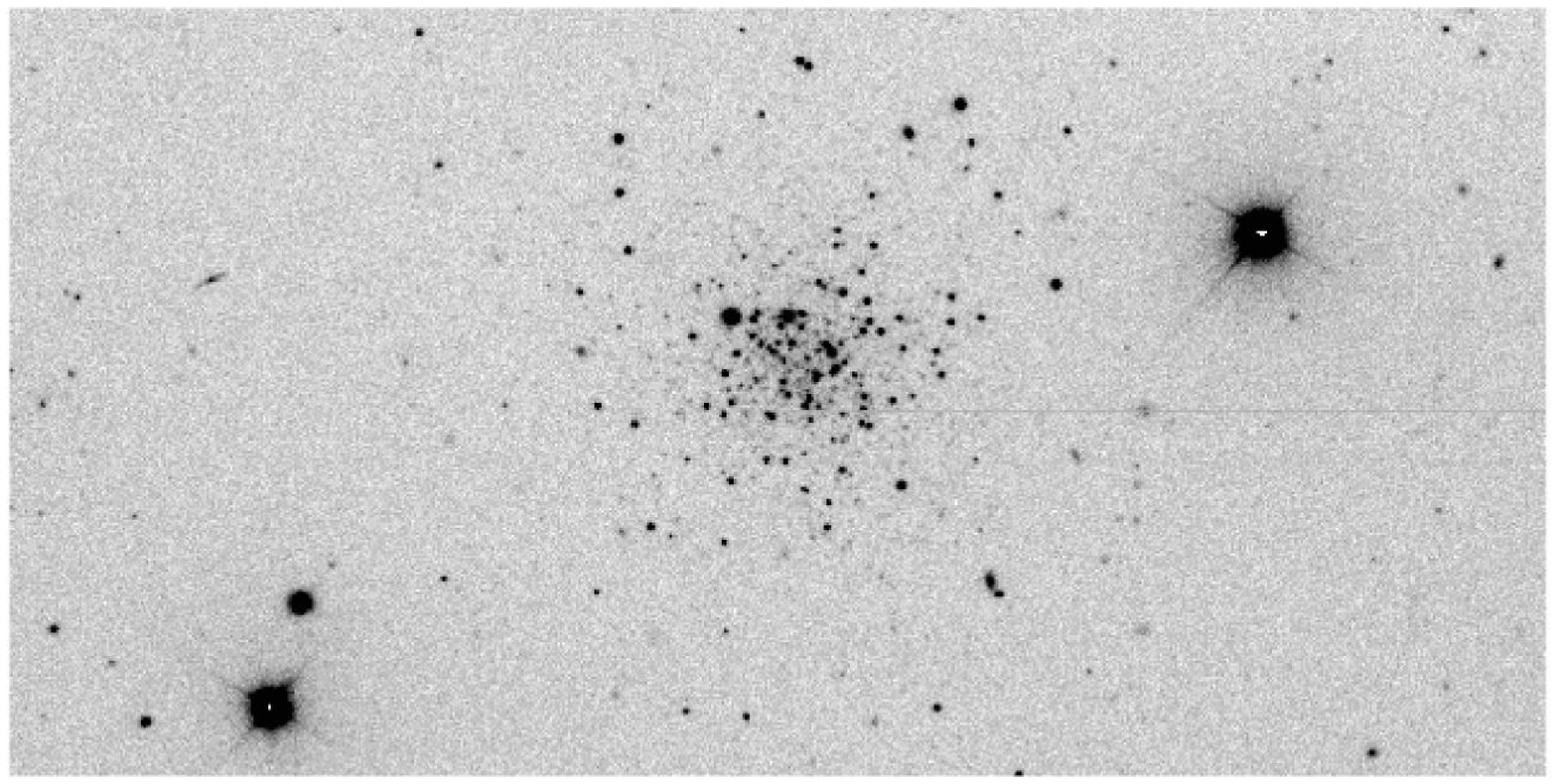,width=8.6cm}
\psfig{figure=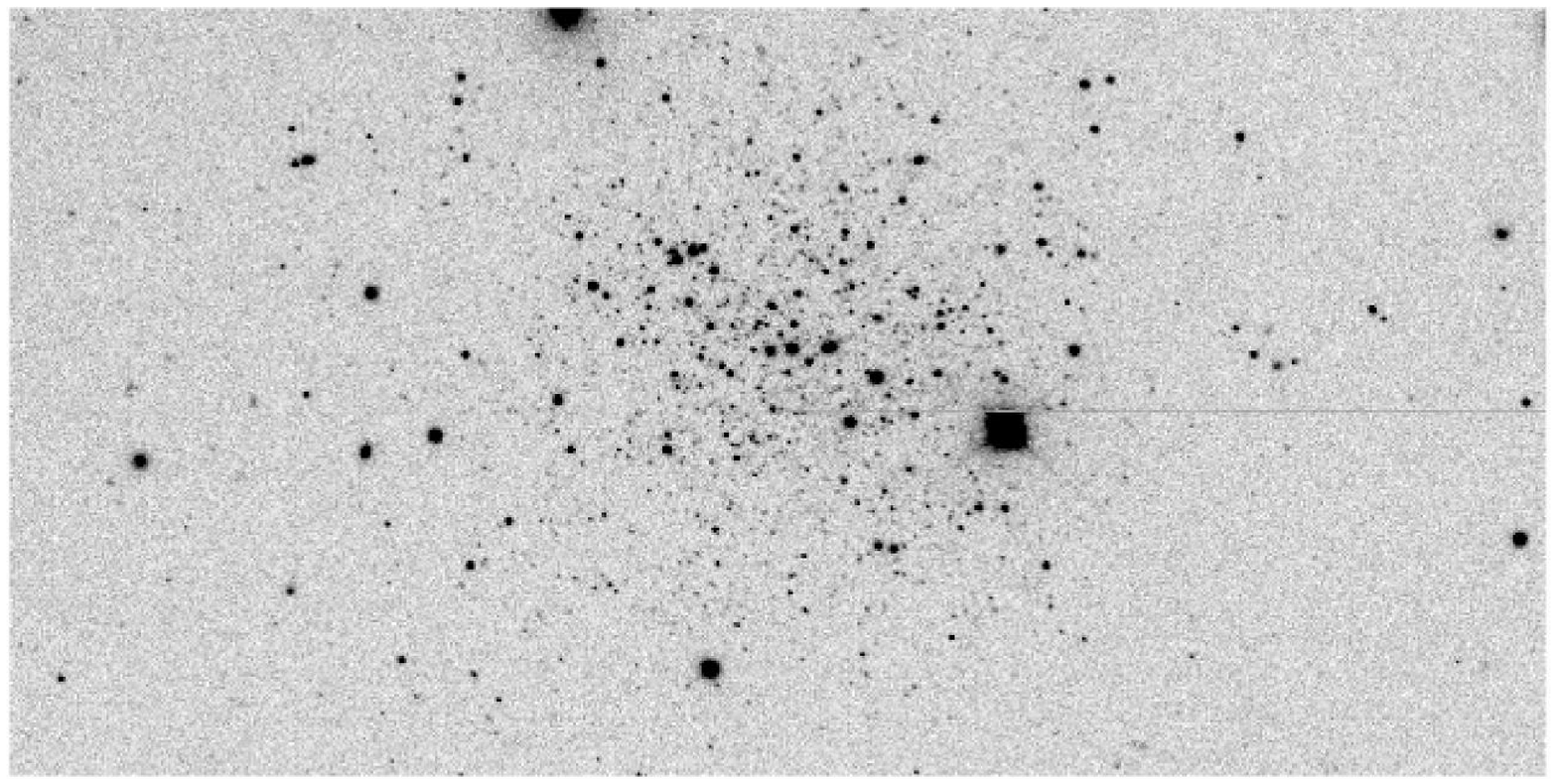,width=8.6cm}
\psfig{figure=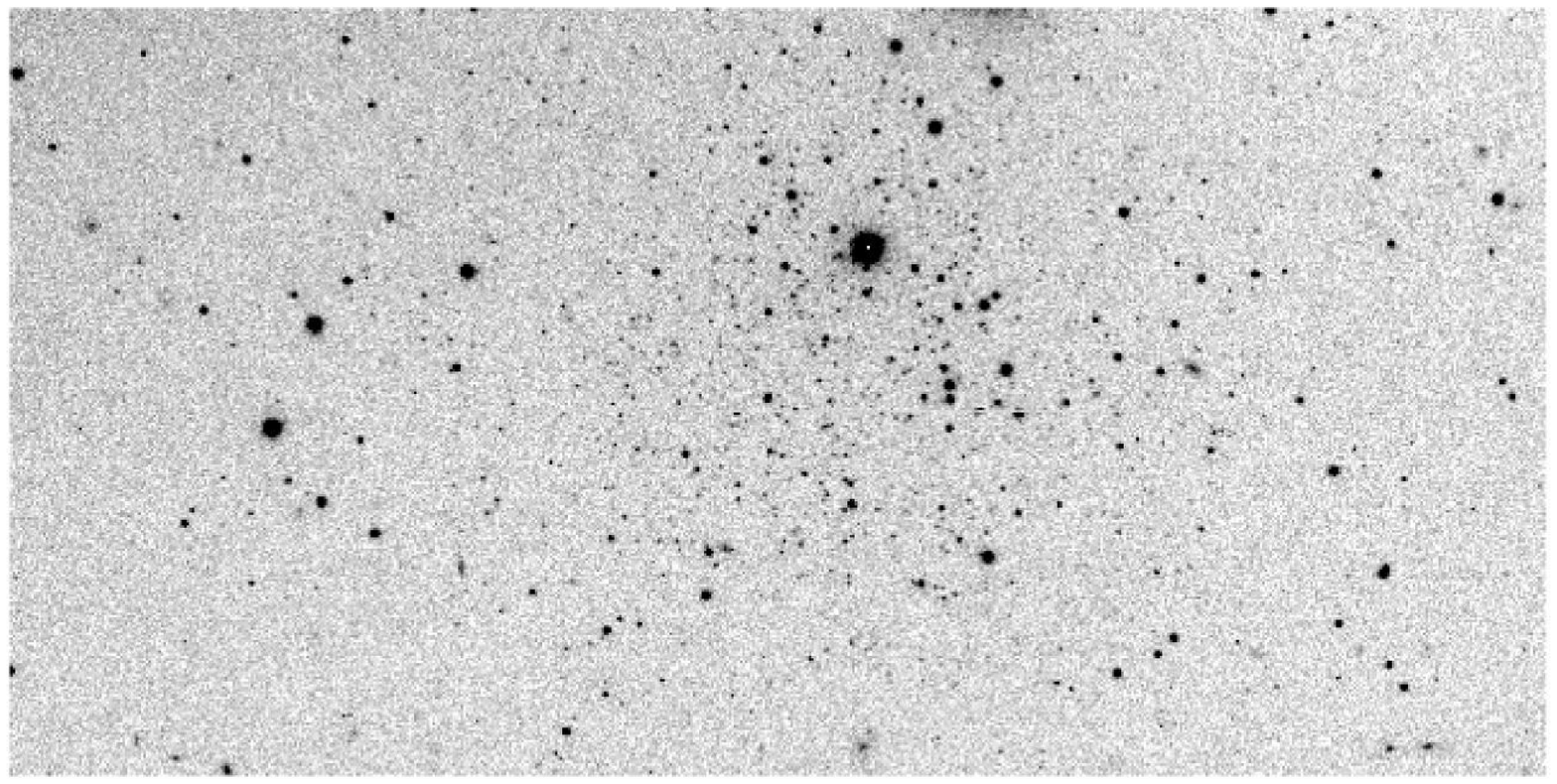,width=8.6cm}
\vspace{0.0cm}
\caption{\label{fig1} Excerpts of the combined $V$ images for AM\,1 (top),
Pal\,3 (middle) and Pal\,14 (bottom). $4\arcmin\times2\arcmin$ sections of the
corresponding `master' chips are shown. North is up, East is left.
}
\end{figure}

\section{Observations and data reduction}\label{obsred}

The observations were performed in February 2005 with the VLT/UT1 at Paranal
(ESO), Chile. The instrument in use was the FORS2 camera with a 4$\times$4\,k
MIT CCD array attached. The data were read out with a 2$\times$2 binning,
resulting in a spatial scale of 0.25\arcsec/pixel and a field of view of
$6\farcm8\times6\farcm 8$.

Each of the three globular clusters AM\,1, Pal\,3 and Pal\,14 was observed
through the Johnson $B$ and $V$ filters under mostly photometric and very
good seeing conditions at two slightly different positions.
Once, centering the cluster on the `master' 4$\times$2\,k chip (also called
``chip1''), and once on the `slave' 4$\times$2\,k chip
($=$``chip2'') of the 1$\times$2 CCD array. Excerpts of the $V$ exposures
on the `master' chip are shown in Fig.~\ref{fig1}.
The exposure times were chosen such that the brightest cluster red giants
were not saturated. Three dithered integrations per filter and field were 
taken. For an observation log see Table~\ref{tab1}.

The CCD frames were processed with standard {\sc IRAF} routines. After bias 
subtraction and flatfield correction, the pixel shifts of the dithered 
images have been determined. Then the shift corrected frames have been
averaged using a clipping algorithm to exclude cosmics. 

\section{Photometry and astrometry}\label{photast}

\subsection{Photometry}\label{phot}

The instrumental
magnitudes of the stars were derived from PSF (point spread function)
photometry using {\sc DAOPHOT II} (Stetson \cite{stet87}, \cite{stet92}).
For the comparison with the standard stars, aperture--PSF shifts have been
determined in all fields and filters. The amplitude of this shift ranges
between 0.20 and 0.44 mag, its uncertainty is of the order of 0.02 mag.

\begin{figure}
\psfig{figure=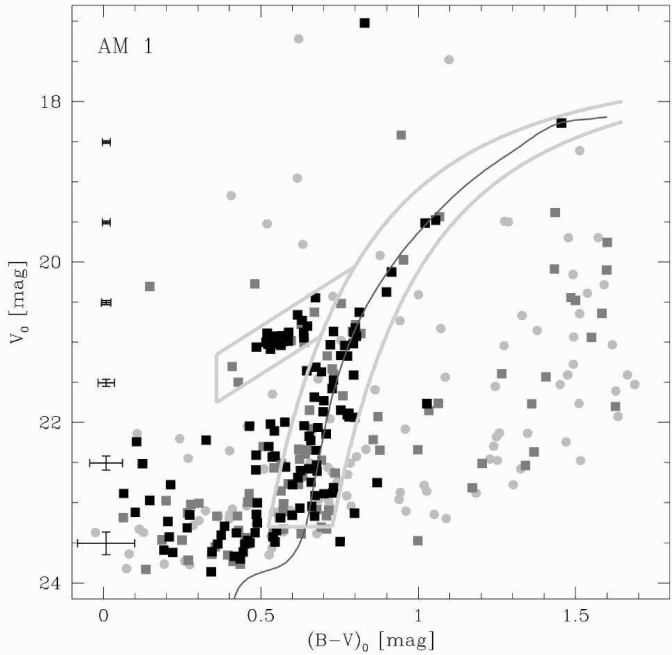,height=8.6cm,width=8.6cm
,bbllx=12mm,bblly=16mm,bburx=190mm,bbury=195mm}
\vspace{0.4cm}
\caption{\label{fig2} Colour magnitude diagram of stars around AM\,1. The
reddening is negligible: $E_{B-V}=0.0$ mag. Dark grey squares are those objects 
that are located within five half-light radii from the centre of the cluster,
black squares those within one half-light radius (see Fig.~\ref{fig3}). Typical 
photometric errors are shown on the left. The selection areas for
probable member stars are marked (red giant branch and horizontal branch).
Overplotted is a Yonsei-Yale isochrone (Kim et al. \cite{kimy02}) with the 
following parameters: age $=11$ Gyr, [Fe/H] $=-1.4$ dex, [$\alpha$/Fe] $=0.3$ 
dex. The isochrone has been shifted with $(m-M)_{V,0}=20.45$ mag and 
$(B-V)_{\rm corr.}=-0.01$ mag (see text for further details). 
}
\end{figure}

\begin{figure}
\psfig{figure=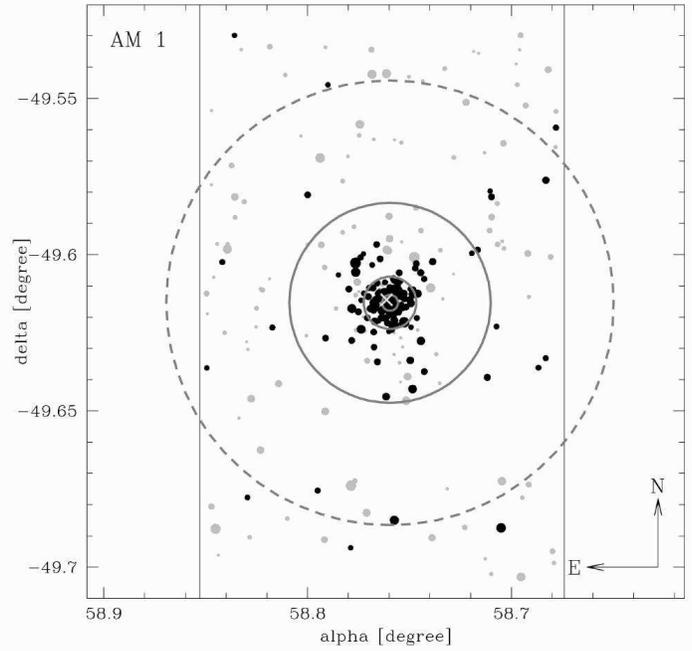,height=8.6cm,width=8.6cm
,bbllx=12mm,bblly=16mm,bburx=190mm,bbury=195mm}
\vspace{0.4cm}
\caption{\label{fig3} Coordinates of stars around AM\,1. The symbol size 
corresponds to the luminosity of an objects: the larger the dot the brighter 
the star. Black dots are probable member stars according to their position in
the CMD (see highlighted areas in Fig.~\ref{fig2}). The solid circles are from 
inside out the core radius, the half-light radius and the tidal radius (Harris
\cite{harr96}). The dashed circle is the tidal radius as calculated by 
Baumgardt et al.  (\cite{baum05}). All circles are centred on the newly 
determined central coordinates of AM\,1 (see Table~\ref{tab3}). The cross marks
the central coordinate from the list of Harris (Harris \cite{harr96}). The
vertical lines indicate the borders of our field-of-view.
}
\end{figure}

In the first night, Landolt standard stars have been observed. The coefficients
of the calibration equation were taken from the ESO quality control program
(see http://www.eso.org/observing/dfo/quality/index\_fors2.html). An overview
of the calibration coefficients is given in Table~\ref{tab2}.

The calibrated magnitudes of stars in the overlapping regions of the North and
South fields have been compared with each other. Shifts of the order of
$\pm0.02$ mag have been found between the data sets. This is consistent with
the uncertainty in the aperture--PSF shift, but could also have been caused by
the not totally photometric conditions during the observing run. Thus, the data
set which was observed under better photometric conditions has been taken
as reference system
for each cluster, and the magnitudes of the other set have been shifted to
this system. The magnitudes of stars that were measured in both data sets
then have been averaged for the final catalog.

After the photometric reduction, calibration of the magnitudes, and combining
of the data sets, the average photometric errors for red giants at $V = 23$
were of the order of 0.08 in $V$ and 0.12 mag in $B$.

\begin{figure}
\psfig{figure=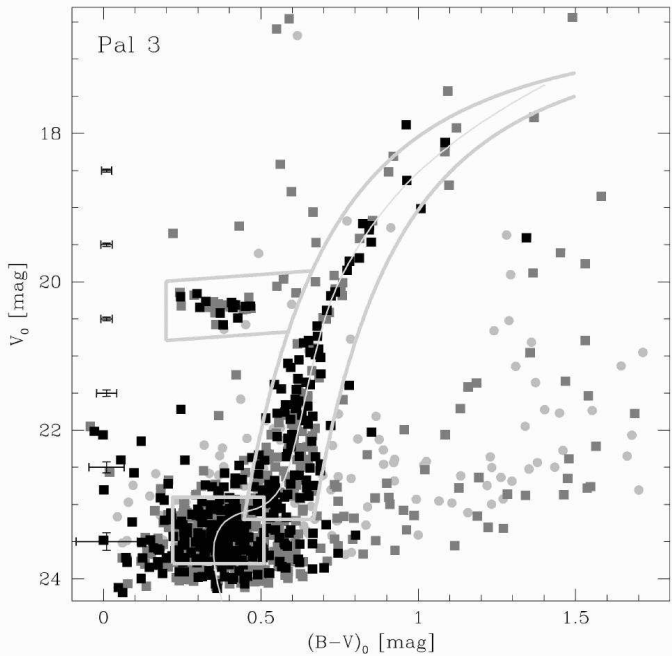,height=8.6cm,width=8.6cm
,bbllx=12mm,bblly=16mm,bburx=190mm,bbury=195mm}
\vspace{0.4cm}
\caption{\label{fig4} Reddening corrected CMD of stars 
around Pal\,3 ($E_{B-V}=0.040$ and $A_V=0.124$ mag). Symbols and areas as in 
Fig.~\ref{fig2}. Additionally, the area of main sequence turnoff stars is 
marked. The isochrone parameters are: age $=10$ Gyr, [Fe/H] $=-1.7$ dex,
[$\alpha$/Fe] $=0.3$ dex, $(m-M)_{V,0}=19.95$ mag, and $(B-V)_{\rm corr.}=-0.02$
mag.
}
\end{figure}

\begin{figure}
\psfig{figure=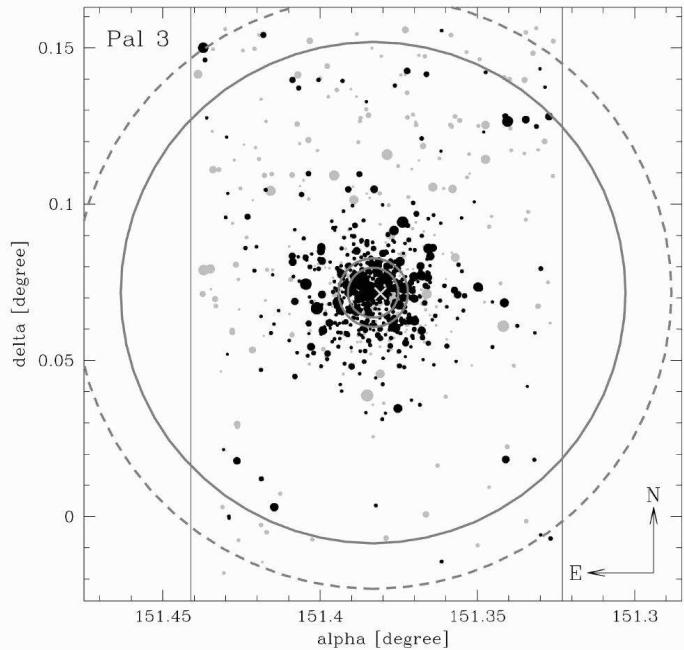,height=8.6cm,width=8.6cm
,bbllx=12mm,bblly=16mm,bburx=190mm,bbury=195mm}
\vspace{0.4cm}
\caption{\label{fig5} Coordinates of stars around Pal\,3. For the explication
of symbols, see Fig.~\ref{fig3}.
}
\end{figure}

\subsection{Astrometry}\label{ast}

The positions of the cluster stars in terms of right ascension and declination
(epoch J2000) were determined by deriving an astrometric solution between
pixel and RA/Dec coordinates for stars of
the US Naval Observatory (USNO) in the field-of-view. Each CCD chip has been
considered separately. Typically, between 60 and 80 USNO catalog stars were
identified in chip1, and 40-60 in chip2. The {\sc IRAF} task {\sc GEOMAP} was
used to find the astrometric solution. A `general' fit with full cross-terms
of second order turned out to give the best results. The rms of the fit in
right ascension is $0\farcs17$ for AM\,1 and Pal\,14, and $0\farcs24$ for
Pal\,3. The accuracy in declination is $0\farcs14$, $0\farcs17$, and
$0\farcs19$ for AM\,1, Pal\,14, and Pal\,3, respectively.

The coordinates of the member stars are listed in the Tables~\ref{tab5},
\ref{tab6}, and \ref{tab7}. The derived coordinates also were used for the
graphs of Figs.\ref{fig3}, \ref{fig5}, and \ref{fig7}.

\section{Distance, reddening, age and metallicity}\label{cmdparam}

The calibrated and de-reddened magnitudes are presented in the
colour-magnitude diagrams (CMDs, see Figs.~\ref{fig2}, \ref{fig4}, and
\ref{fig6}). The foreground reddening has been corrected with the $E_{B-V}$
values given in the Harris list (Harris \cite{harr96}), and $A_V = 3.1
E_{B-V}$. Only stars with photometric errors less than 0.15 mag in $V$ and $B$
and {\sc ALLSTAR} {\it sharp} and {\it chi} values in the ranges
$-0.7<$\,{\it sharp}\,$<1.0$ and {\it chi}\,$<3.0$ are shown. The red giant 
branch (RGB) and horizontal branch (HB) are clearly visible in all clusters. 
For Pal\,3, even the main sequence turnoff (MSTO) region is well definded.

The distances to the clusters are derived from the apparent magnitude of the
horizontal branch. We selected only those HB stars which are located within
five half-light radii from the clusters' centres and which lie on the
horizontal part of the HB at $(B-V)_0 \sim 0.5$ (to avoid AGB and extended HB
stars). The average HB magnitudes in $V$ and the corresponding distance moduli
are given in Table~\ref{tab3}. They
agree within $\pm0.03$ mag with the values listed by Harris (\cite{harr96}).

In order to estimate the age and metallicity of the clusters we fitted the
location and shape of the RGBs by appropriate isochrones. The Yonsei-Yale
isochrone set by Kim et al. (\cite{kimy02}) was used for that. An
alpha-abundance of [$\alpha$/Fe]$=0.3$ dex was adopted. The best-fitting
isochrones are shown in Figs.~\ref{fig2}, \ref{fig4}, and \ref{fig6}. They
have ages between 10 and 11 Gyr, and metallicities in the range $-1.7$ to
$-1.4$ dex. For Pal\,3 and Pal\,14 the best-fitting metallicity value agrees
well with the spectroscopic values in the Harris list (see Table~\ref{tab3}).
For AM\,1, no good isochrone fit with a metallicity of $-1.8$ dex could be
found. Instead, a metallicity of $-1.4$ dex represents the shape of the RGB
well. Either, the spectroscopic abundance is not accurate enough and this
cluster indeed has a higher metallicity or, the isochrone fit is not
constrained well enough. Besides the selection of the best-fitting ages and
metallicities, the isochrones had to be slightly shifted in $V$ as well as
in $(B-V)$ to match the exact location of the RGB. These corrections are
given in the captions of the CMD figures, $(m-M)_{V,0}$ and $(B-V)_{\rm corr.}$,
and in Table~\ref{tab3} (with $E_{B-V,\rm this\,work} = E_{B-V,\rm
Harris}+(B-V)_{\rm corr.}$). These corrections are of the order of the
photometric errors and within the uncertainties of not perfectly photometric 
conditions.

\subsection{Member selection}\label{membsel}

\begin{figure}
\psfig{figure=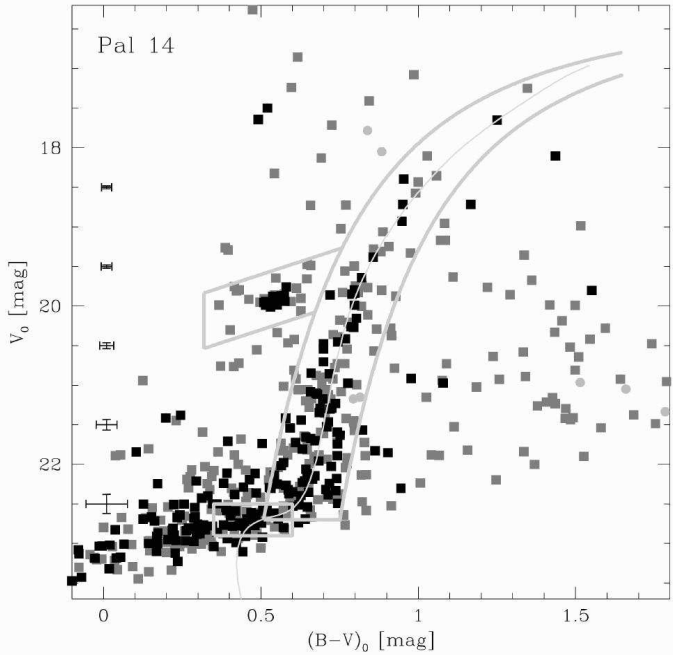,height=8.6cm,width=8.6cm
,bbllx=12mm,bblly=16mm,bburx=190mm,bbury=195mm}
\vspace{0.4cm}
\caption{\label{fig6} Reddening corrected CMD of stars
around Pal\,14 ($E_{B-V}=0.040$ and $A_V=0.124$ mag). Symbols and areas as in
Fig.~\ref{fig2}. The lowest area includes probable subgiant branch stars. The
isochrone parameters are: age $=10$ Gyr, [Fe/H] $=-1.5$ dex,
[$\alpha$/Fe] $=0.3$ dex, $(m-M)_{V,0}=19.42$ mag, and $(B-V)_{\rm corr.}=0.03$
mag.
}
\end{figure}

The main purpose of this paper is to provide a list of bright cluster members
for follow-up spectroscopy. Therefore, probable cluster members have to be
selected. The principle criterium is that member stars should be located close
to the isochrone in the colour-magnitude diagram. A second criterium is their
distance to the cluster centre. In the Figs.~\ref{fig2}, \ref{fig4}, and
\ref{fig6} the regions of probable member stars are high-lighted. The RGB and
HB regions are marked for all clusters. In Pal\,14, an additional region of
subgiant branch (SGB) stars has been selected, in Pal\,3 also the MSTO region.
Furthermore, stars within one and five half-light radii are shown in the graphs
as black and dark grey squares, respectively. The core radius, half-light 
radius and tidal radius (Harris \cite{harr96}, Baumgardt et al. \cite{baum05}) 
are listed
in Table~\ref{tab4} and shown in Figs.~\ref{fig3}, \ref{fig5}, and \ref{fig7}.

In the Appendix, magnitude limited samples of probable member stars
are listed (Tables~\ref{tab5}, \ref{tab6}, and \ref{tab7}).
Only stars within the tidal radius given by Baumgardt et al. (\cite{baum05})
have been considered. The magnitude limits are $V_0 = 23.0$, 22.5, and 22.2
for AM\,1, Pal\,3, and Pal\,14, respectively. Thus, only RGB, HB and AGB stars
are contained in the lists. Note: the full tables only are availble in the
online version of the article.

\section{Structural parameters}\label{struct}

Although the photometry of the observed clusters is not very deep and we
do not sample the main sequence stars, some structural parameters of the
globular clusters can be derived from the selected evolved member stars.

\subsection{Cluster centres}\label{centres}

\begin{figure}
\psfig{figure=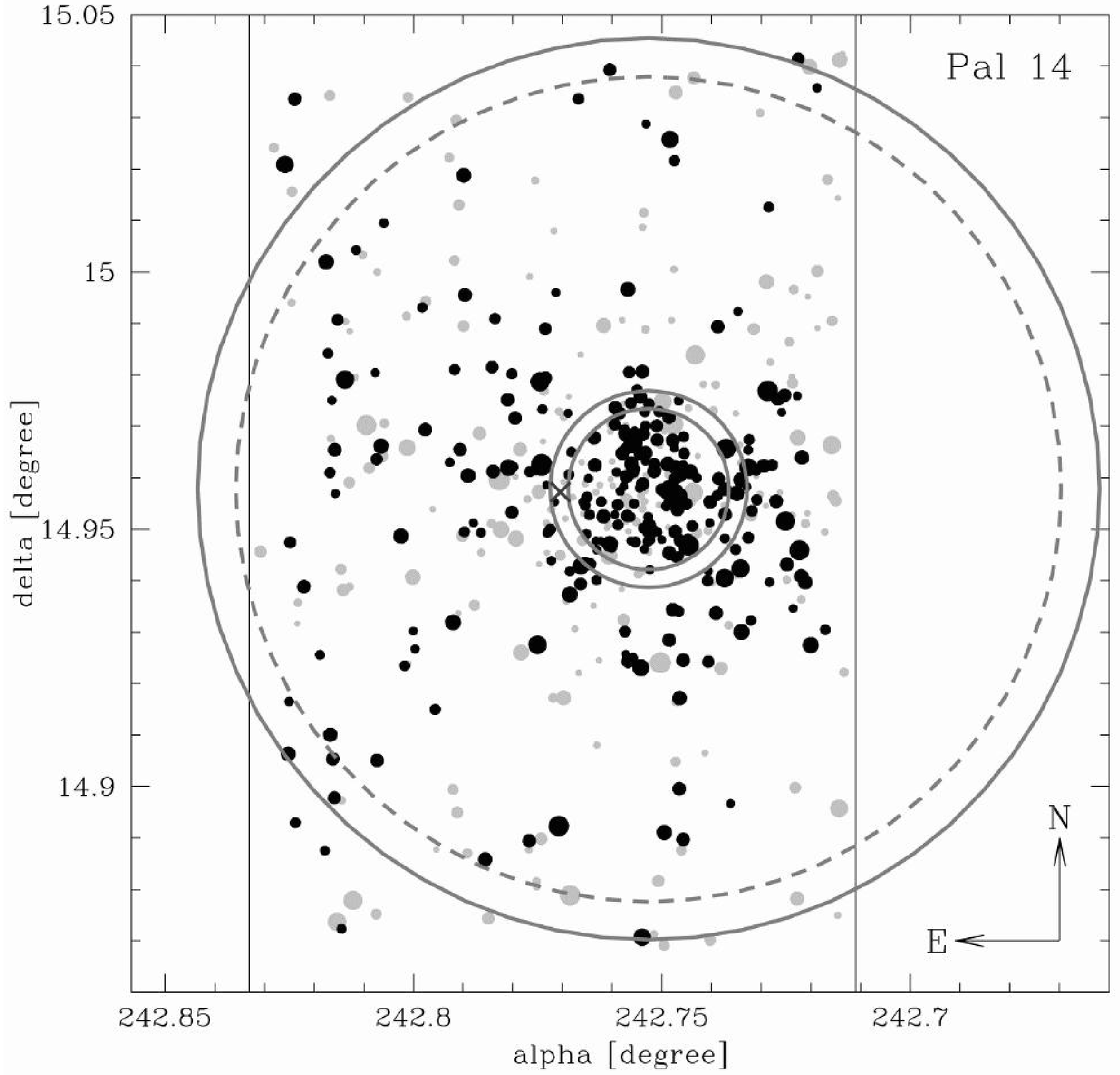,height=8.6cm,width=8.6cm
,bbllx=12mm,bblly=16mm,bburx=190mm,bbury=195mm}
\vspace{0.4cm}
\caption{\label{fig7} Coordinates of stars around Pal\,14. For the explication
of symbols, see Fig.~\ref{fig3}.
}
\end{figure}

\begin{figure}
\psfig{figure=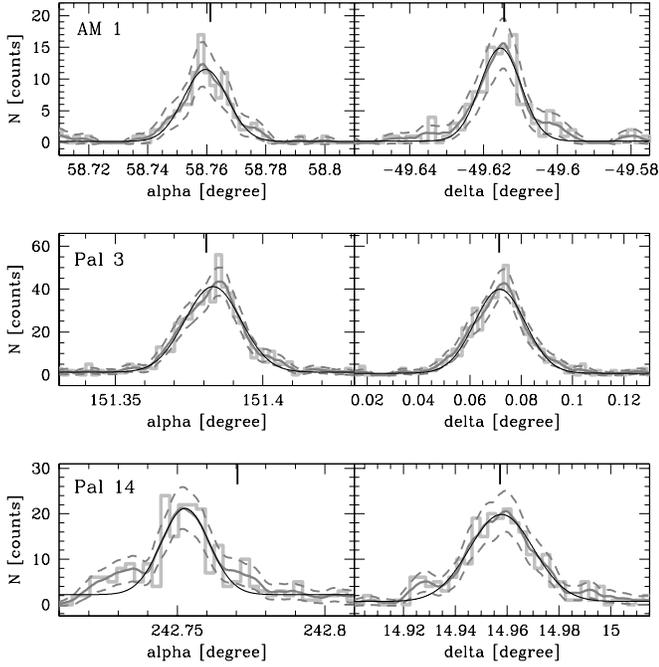,height=8.6cm,width=8.6cm
,bbllx=9mm,bblly=65mm,bburx=195mm,bbury=244mm}
\vspace{0.4cm}
\caption{\label{fig8} Determination of globular cluster centres from the
spatial distribution of selected probable member stars (see the CMDs). 
The grey histograms in the left panels show the distribution in 
West-East-direction, the ones in the right panels the distribution in
South-North-direction. Binning independent representations of the counts 
(Epanechnikov kernel of 0.002 (AM\,1 and Pal\,3) and 0.003 (Pal\,14) degree
width) are overplotted as thin curves, together with their $1\sigma$ 
uncertainty limits (dashed curves). The thick-lined Gaussians are fits to the
binning independent distributions. The bold tickmarks on top of each plot
mark the central positions from the Harris catalog (\cite{harr96}).
}
\end{figure}

The cluster centres were determined by analysing the number distribution
of probable member stars as function of right ascension and declination. 
Probable member stars are those of the high-lighted regions in the CMDs
(see Figs.~\ref{fig2}, \ref{fig4}, and \ref{fig6}). Furthermore, the
member stars have been restricted to an approximatly squared region 
($6.5\arcmin\times6.5\arcmin$) around
their apparent cluster centres. The number distributions of the stars are
shown in Fig.~\ref{fig8} as histograms and as binning independent 
representations
(using an Epanechnikov kernel). The binning independent number distributions 
were fitted by Gaussians. The peak of these Gaussians defines the cluster
centre. In Table~\ref{tab3} the derived values are listed. The accuracy of
the centre determination is about $\pm2\arcsec$. Whereas the declination of the
centres agree very well with those listed in the Harris list, there are some
notable deviations in the right ascension (see also crosses in Figs.~\ref{fig3},
\ref{fig5}, and \ref{fig7}). Especially, the right ascension listed for Pal\,14 
is off by $\sim62\arcsec$ with respect to our value. We are confident
that our values are very accurate and suggest to correct the coordinates in 
the `official' lists (e.g. Harris \cite{harr96}).

\subsection{Angular distribution and ellipticity}

\begin{figure}
\psfig{figure=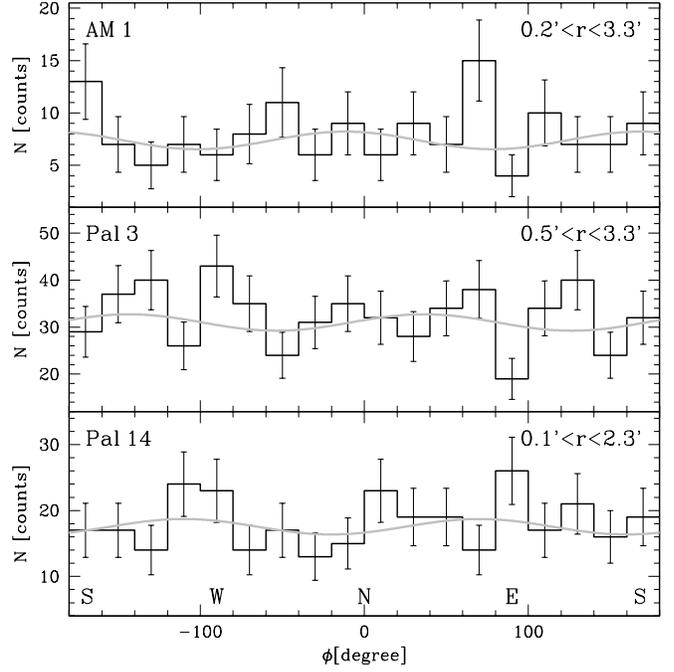,height=8.6cm,width=8.6cm
,bbllx=9mm,bblly=65mm,bburx=195mm,bbury=244mm}
\vspace{0.4cm}
\caption{\label{fig9} The angular distribution of probable member stars in
the indicated radial ranges around the three clusters is shown together with
statistical errorbars. The grey curves
are best-fitting double-cosines, indicating the ellipticity of the clusters.
}
\end{figure}

The ellipticity of the clusters is of interest because rotation or the tidal 
loss of stars in opposite directions might cause a detectable signal.
To investigate this we studied the angular distribution of probable member 
stars including the main sequence turnoff region stars (see 
Fig.~\ref{fig9}).
We choose the radial range to be from outside the core radius
(only for AM\,1 and Pal\,3) to avoid crowding effects to inside the maximum
radius of the observed CCD area to guarantee a homogeneous angular coverage.
The angular distributions (counted in bins of 20 degree) then were fitted by a 
double-cosine function of the form $N = A + B\cdot \cos(2\cdot(\phi+C))$, where
$A$ and $B$ define the ellipticity as follows: $\epsilon = (A-B)/(A+B)$, 
assuming that the ellipticity is expressed as $1-b/a$ ($a$: semi 
major axis, $b$: semi minor axis). The error for the ellipticity
was propagated from the Poisson statistics of the number counts and the rms of
the fit. The results are: $\epsilon_{\rm AM\,1} = 0.20\pm0.26$,
$\epsilon_{\rm Pal\,3} = 0.11\pm0.20$, and $\epsilon_{\rm Pal\,14} = 
0.13\pm0.10$. None of the clusters exhibits a significant ellipticity.
The distribution of their stars is compatible with a homogeneous,
circular distribution within the errors.

\subsection{Surface density profiles}\label{surfdens}

\begin{figure}
\psfig{figure=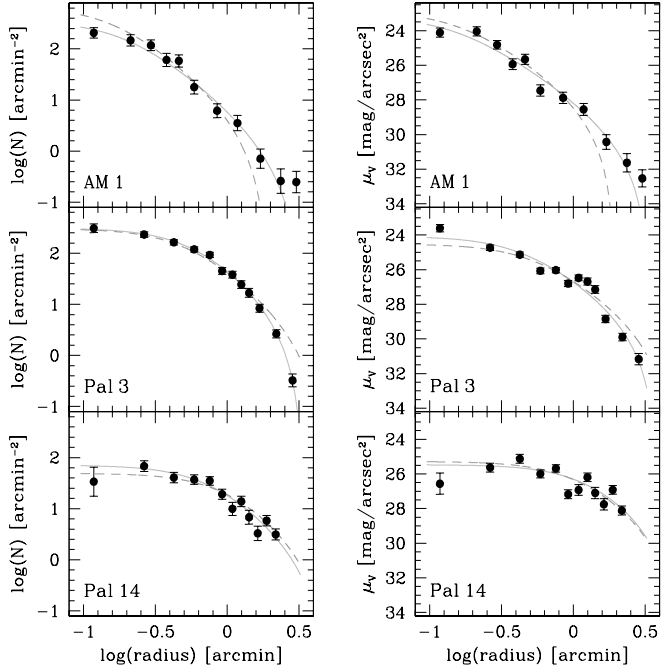,height=8.6cm,width=8.6cm
,bbllx=9mm,bblly=65mm,bburx=195mm,bbury=244mm}
\vspace{0.4cm}
\caption{\label{fig10} The surface number density distribution (left panels)
and the surface brightness distribution (right panels) of probable member stars
is shown for the three clusters. The grey curves are best-fitting King models.
Dashed curves are fits to the distributions when adopting the values for $r_c$
and $r_t$ of Harris' list (\cite{harr96}).
The parameters of both fits are given in Table~\ref{tab4}.
}
\end{figure}

The probable member stars of the three clusters are evolved stars of more or
less the same mass. Since they are the brightest stars of the clusters they
dominate the surface brightness of those clusters. One can derive the
structural parameters of the clusters by fitting density profiles to either
the number density distribution or the surface brightness distribution of
the stars. This is shown in Fig.~\ref{fig10}. The number density counts have 
been corrected for background/foreground field star contamination by estimating
the field star number counts from 0.5 arcmin wide strips in the northern and 
southern edge of the observed CCD areas. The number density of 
contaminating field stars in the selected CMD areas is 0.15, 0.77, and 0.62 
objects/arcmin$^{-2}$ for AM\,1, Pal\,3, and Pal\,14, respectively.
The core radius, the tidal radius, and the central surface brightness of the
clusters were determined from the best fitting King (King \cite{king62})
profiles. For Pal\,14, the tidal radius had to be fixed to 5 arcmin (the mean
value between Harris \cite{harr96} and Baumgardt et al. \cite{baum05}) since it
could not be constrained by the fit. The half-light radii were derived from a
curve-of-growth analysis of the integrated light of the member stars. The
results of our analyses are given in Table~\ref{tab4}. Typical errors are of
the order $0\farcm06$ (AM\,1, Pal\,3) and $0\farcm20$ (Pal\,14) for $r_c$,
$0\farcm04$ (Pal\,3) and $0\farcm19$ (AM\,1) for $r_t$, while for $r_h$ they
are $\sim0\farcm02$ respectively. The errors for the central surface brightness
are between 0.25 and 0.40 mag.

Not all our derived values agree with the ones listed by Harris (\cite{harr96})
and/or Baumgardt et al. (\cite{baum05}). For AM\,1, we find a slightly smaller
half-light radius, but a significantly larger tidal radius than listed by
Harris. A small tidal radius does not fit the profile. Also the the central
surface brightness is much higher than Harris' value. In the contrary, for
Pal\,3, we derive a slightly larger half-light radius, but a smaller tidal
radius. Also for Pal\,14, a larger half-light radius fits the profile best.

Note however that all our values are valid for the selected probable member
stars. Adding the fainter, lower mass main sequence stars would raise the
central surface brightness a little bit. One might think that including the
faint stars would increase the characteristic radii since these stars might be
more widely distributed than the more massive, evolved stars due to the
dynamical evolution of the cluster. However, the half-mass relaxation times
of the clusters are between 5.5 and 9.1 Gyr, close to a Hubble time. This
suggests that dynamical evolution does not play a major role in the shaping of
the clusters.

\section{Discussion and Summary}\label{summcon}

The outer halo globular clusters AM\,1, Pal\,3, and Pal\,14 were imaged
with FORS2 on the VLT under good seeing and mostly photometric conditions.
Accurate PSF photometry in Johnson $BV$ of stars brighter than $V=24$ mag
was performed in $6\farcm8\times10\farcm8$ large fields around these clusters.
The absolute position of the stars was determined with an accuracy of
$<$0.2$\arcsec$ by applying a general astrometric solution derived from stars
of the USNO catalog in the observed fields. This positional accuracy is
sufficient to be used in input catalogs for most of the available multi-object
spectrographs (using either fibres or slit masks).

Probable member stars of the clusters were selected according to their position
in the colour-magnitude diagram and their distance to the cluster centres.
The positions and photometric properties ($V$ magnitude, $(B-V)$ colour) of all
probable member stars within the tidal radius (as calculated by Baumgardt et
al.~\cite{baum05}) and brighter than $V_0=23.0,22.5,22.2$ mag for AM\,1,
Pal\,3, and Pal\,14, respectively, are presented in the tables of the appendix.
Although challenging, one can reach with modern high resolution spectrographs,
like UVES at the VLT, a signal-to-noise above 10 for stars as faint as $V=20.5$
mag within one observing night. This limit would mean a sample size of 9, 48,
and 72 probable member stars for AM\,1, Pal\,3, and Pal\,14, respectively.
Except for AM\,1, these sample sizes are sufficient to derive an accurate
radial velocity dispersion of the clusters.

The colour-magnitude diagrams have been used to estimate -- by isochrone
fitting -- distances, reddenings, ages and metallicities to the three clusters.
Our derived values for the horizontal branch magnitude $V_{\rm HB}$, reddening
and distance modulus $(m-M)_V$ generally agree within $\pm0.03$ mag with the
values given in the literature (AM\,1: Madore \& Freedman \cite{mado89};
Pal\,3: Stetson et al. \cite{stet99}; Pal\,14: Holland \& Harris
\cite{holl92}). Only for Pal\,14 we get a slightly better isochrone fit
when adopting a 0.05 mag shorter distance modulus and a 0.03 mag higher
reddening (see Table~\ref{tab3}).
Since the turnoff regions are not very well constrained in the CMDs of the
three clusters, age estimates have to be taken with caution. However, we found
that isochrones with ages of 10-11 Gyr fit the turnoff-SGB-RGB transition
region better than those of older ages. Our age estimates are consistent with
the idea that the three clusters belong to the 'young halo' GCs which are, on
average, 1-2 Gyr younger than the old 'inner halo' GCs (e.g. Mackey \& Gilmore
\cite{mack04}). The metallicity of the chosen isochrones is mainly constrained
by the shape of the red giant branch. The adopted values for Pal\,3 ($-1.7$
dex) and Pal\,14 ($-1.5$ dex) agree well with those given in the literature:
Pal\,3: $-1.78$ (Zinn \cite{zinn85}), $-1.57\pm0.3$ (Armandroff et al.
\cite{arma92}), $-1.70\pm0.15$ (Ortolani \& Gratton \cite{orto89}); Pal\,14:
$-1.47\pm0.3$ (Zinn \cite{zinn85}), $-1.60\pm0.18$ (Armandroff et al.
\cite{arma92}). The isochrone of AM\,1, however, fits the RGB better with a
higher metallicity ($-1.4$ dex) than derived in previous work: $-1.69$ (Zinn
\cite{zinn85}), $-1.7\pm0.2$ (Suntzeff et al. \cite{sunt85}). New spectroscopic
metallicities are needed to decide whether AM\,1 really is more metal-rich
than thought before.

\begin{table*}[t!]
\caption{\label{tab3} Basic photometric parameters of AM\,1, Pal\,3, and 
Pal\,14. The second row for each cluster gives the values from Harris' list 
(\cite{harr96}). $E_{B-V,\rm S98}$ is the reddening value by Schlegel et al.
(\cite{schl98}). The magnitude of the horizontal branch $V_{\rm HB}$ and the 
total apparent $V$ magnitude $V_{\rm tot}$ are not corrected for extinction. 
The first distance modulus is based on the derivation of the absolute magnitude 
of the HB using $M_{V,\rm HB} = 0.15$[Fe/H]$+0.8$, the second one on the 
isochrone fitting. The distance to the Sun $R_{\rm Sun}$ was calculated from
the first distance modulus after correction for extinction with 
$A_V = 3.1E_{B-V,\rm S98}$.}
\begin{tabular}{lcc@{\hspace{2mm}}c@{\hspace{2mm}}cc@{\hspace{2mm}}c@{\hspace{2mm}}c@{\hspace{2mm}}cccc}
\hline
 & & & & & & & & & & & \\[-3mm]
Name & RA(2000) & Dec(2000) & $E_{B-V,\rm S98}$ & $E_{B-V}$ & $V_{\rm HB}$ & 
$(m-M)_{V,\rm HB}$ & $(m-M)_{V,\rm iso}$ & [Fe/H] & $R_{\rm Sun}$ & 
$V_{\rm tot}$$^a$ & $M_V$ \\
 & [h:m:s] & [$^\circ$:$\arcmin$:$\arcsec$] & [mag] & [mag] & [mag] & [mag] & 
 [mag] & [dex] & [kpc] & [mag] & [mag] \\
\hline
 & & & & & & & & & & & \\[-3mm]
AM\,1   & 03:55:02.3 & $-$49:36:55 & 0.01 & ... & 20.98 & 20.39 & 20.45 & 
$-$1.40 & 118.0 & 15.84 & $-$4.55 \\
        & 03:55:02.7 & $-$49:36:52 & ...  & 0.00 & 20.96 & 20.43 &  ...  & 
$-$1.80 & 121.9 & 15.72 & $-$4.71 \\
 & & & & & & & & & & & \\[-3mm]
Pal\,3  & 10:05:31.9 &    00:04:18 & 0.04 & 0.02 & 20.48 & 19.94 & 19.95 & 
$-$1.70 & 91.9 & 14.91 & $-$5.03 \\
        & 10:05:31.4 &    00:04:17 & ...  & 0.04 & 20.51 & 19.96 &  ...  & 
$-$1.66 & 92.7 & 14.26 & $-$5.70 \\
 & & & & & & & & & & & \\[-3mm]
Pal\,14 & 16:11:00.6 &    14:57:28 & 0.03 & 0.07 & 20.03 & 19.46 & 19.42 & 
$-$1.50 & 74.7 & 14.68 & $-$4.78 \\
        & 16:11:04.9 &    14:57:29 & ...  & 0.04 & 20.04 & 19.47 &  ...  & 
$-$1.52 & 73.9 & 14.74 & $-$4.73 \\
\hline
\end{tabular}
\vskip0.1cm
$^a$ The total magnitude in the first row for each cluster is based on the 
integrated luminosity of probable member stars down to the magnitude 
\hspace*{0.2cm} limits of the individual data sets. It therefore excludes most 
of the bright foreground stars, but also the fainter main sequence stars.
\end{table*}

\begin{table*}[t!]
\caption{\label{tab4} Basic structural parameters of AM\,1, Pal\,3, and 
Pal\,14. The second row for each cluster gives the values from Harris' list
(\cite{harr96}). The last three columns are physical scales based on the 
distances in Table~\ref{tab3}.}
\begin{tabular}{lcccccccc}
\hline
 & & & & & & & & \\[-3mm]
Name & $\mu_{V,\rm central}$ & $r_c$$^a$ & $r_h$ & $r_t$$^a$ & c$^{a,b}$ &
$r_c$ & $r_h$ & $r_t$ \\
 & [mag/arcsec$^2$] & [$\arcmin$] & [$\arcmin$] & [$\arcmin$] & & [pc] & [pc] &
[pc] \\
\hline
 & & & & & & & & \\[-3mm]
AM\,1   & 23.11 & 0.18 (0.14) & 0.32 & 3.24 (3.29) & 1.26 (1.37) & 6.2 (4.8) &
11.0 & 111.3 (113.0) \\
        & 23.86 & 0.15        & 0.50 & 1.92 (4.27) & 1.12        & 5.3       &
17.7 &  68.1 (151.6) \\
 & & & & & & & & \\[-3mm]
Pal\,3  & 23.84 & 0.54 (0.40) & 0.72 & 3.34 (3.72) & 0.79 (0.97) & 14.4 (10.7) &
19.3 &  89.3 (99.5) \\
        & 23.08 & 0.48        & 0.66 & 4.81 (5.69) & 1.00        & 12.9      &
17.8 & 129.8 (153.5) \\
 & & & & & & & & \\[-3mm]
Pal\,14 & 24.92 & 0.70 (1.14) & 1.28 & [5.00]$^c$  & 0.85 (0.59) & 15.2 (24.8) &
27.8 & [108.7]$^c$ \\
        & 25.55 & 0.94        & 1.15 & 5.26 (4.81) & 0.75        & 20.21     &
24.7 & 113.1 (103.4) \\
\hline
\end{tabular}
\vskip0.1cm
$^a$In brackets the values of the fit to the surface brightness profile are
given (first row for each cluster). The tidal radii in brackets of the
second \hspace*{0.1cm} row for each cluster are the values from Baumgardt et
al. (\cite{baum05})\\
$^b$ concentration parameter: $c = \log(r_t/r_c)$\\
$^c$ this value was not derived but adopted
\end{table*}

From the spatial distribution of probable member stars in AM\,1, Pal\,3, and 
Pal\,14 structural parameters of the clusters have been determined and
compared to the values given in Harris' list (\cite{harr96}) and Baumgardt et
al. (\cite{baum05}). Whereas the derived declinations of all clusters agree 
with the literature values within $\pm3\arcsec$, the right ascensions show
larger deviations: $4\arcsec$ for AM\,1, $7\arcsec$ for Pal\,3, and most
significantly $62\arcsec$ for Pal\,14. The deviating right ascension of Pal\,14 
goes back 
to the list of Webbink (\cite{webb85}), and then was taken over into the lists
of Djorgovski \& Meylan (\cite{djor93}) and Harris (\cite{harr96}).

Concerning ellipticities, none of the clusters shows a measurable signal.
They all seem to be relaxed, roundish halo clusters. Note however that a wide 
field study of main sequence stars around Pal\,3 revealed a weak
signature of extra tidal member stars between one and four tidal radii (Sohn
et al. \cite{sohn03}). 

The half-light radii of the clusters were estimated from curve-of-growth
analyses of the integrated light of the probable member stars. Our derived 
values deviate by $\pm0.2\arcmin$ from those given in Harris' list which
are based on CCD work by Trager er al. (\cite{trag93}) and van den Bergh et al.
(\cite{vdbe91}). Biases in the different selection of member stars and the 
accidental inclusion of foreground stars might explain these differences.
The core and tidal radii of the three clusters were derived from King profile
fits to their surface number density as well as surface brightness profiles.
For AM\,1 and Pal\,3 both radii could be determined with stable solutions.
For Pal\,14, the spatial extension of the member stars was not
sufficient enough to get a stable fit for the tidal radius. It was fixed to
$5\arcmin$, the mean value of Harris (\cite{harr96}) and Baumgardt et al.
(\cite{baum05}). The core and tidal radius of Pal\,14 in Harris' list is based
on the photographic study by Harris \& van den Bergh (\cite{harr84}), the ones
of AM\,1 and Pal\,3 on the CCD work by Trager er al. (\cite{trag93}).
The main differences of our findings to previous literature values are that
AM\,1 definitly has a larger tidal radius than claimed by Trager et al. 
(\cite{trag93}), but smaller than calculated by Baumgardt et al. 
(\cite{baum05}). On the contrary, the tidal
radius of Pal\,3 seems to be smaller than claimed by the two authors. The core
radii of all clusters agree more or less with those given in Harris' list.
The three clusters have in common that their central concentration parameter
is around 1.0, very low in comparison with ordinary inner halo globular 
clusters.\\

This paper provides the basis for further investigations of evolved stars
in AM\,1, Pal\,3, and Pal\,14.
Future follow-up spectroscopic observations of these cluster will be very
exciting. The measurements of a low (Newtonian) velocity dispersion would mean
that MOND in its present form is in severe trouble and that globular clusters
do not possess dark matter. In contrast, a high velocity dispersion would
either favour MOND or could be a hint to the existence of dark matter in
globular clusters, a highly interesting hypothesis on its own right.

\acknowledgements
I wish to thank Holger Baumgardt and Klaas de\,Boer for very helpful comments,
and the anonymous referee for useful suggestions which improved the paper.
This project was supported by the DFG project HI 855/2.

\appendix
\section{Lists of probable member stars}

The following tables contain the lists of probable member stars of AM\,1,
Pal\,3, and Pal\,14. They are ordered with increasing $V$ magnitude. The 
columns are as follows:\\

\noindent
{\bf Column 1.} Identification number of the object, following the order of
increasing $V$ magnitude.

\noindent
{\bf Column 2.} Right ascension for the epoch 2000 in hours, minutes and
seconds ($^h$,$^m$,$^s$).

\noindent
{\bf Column 3.} Declination (2000) in degrees, minutes and seconds 
($\degr,\arcmin,\arcsec$).

The positions of all objects were determined relative to positions in the USNO 
catalog. The positional accuracy of the calculated coordinates is in all 
fields better than 0.2\arcsec.

\noindent
{\bf Column 4.} $V$ apparent magnitude as determined by PSF photometry under
{\sc DAOPHOT II}.

\noindent
{\bf Column 5.} $B-V$ colours from PSF photometry.

\noindent
{\bf Column 6.} Photometric error in $V$ ({\sc ALLSTAR} output).

\noindent
{\bf Column 7.} Photometric error in $B$ ({\sc ALLSTAR} output).

\noindent
{\bf Column 8.} Projected distance to cluster centre (see Table~\ref{tab3})
in arcminutes.

\noindent
{\bf Column 9.} `Type' describes to which part of the CMD the star probably 
belongs: RGB = red giant branch, HB = horizontal branch, AGB = asymptotic
giant branch.\\

\begin{table}[t!]
\caption{\label{tab5} List of probable member stars of AM\,1, ordered with
increasing $V$ magnitude.}
{\tiny
\begin{tabular}{r@{\hspace{3mm}}c@{\hspace{3mm}}cc@{\hspace{3mm}}c@{\hspace{3mm}
}c@{\hspace{3mm}}ccc}
\hline
 & & & & & & & & \\[-1mm]
Id & $\alpha$(2000) & $\delta$(2000) & $V$ & $B-V$ & $\sigma_V$ & $\sigma_B$ &
 radius & Type \\
 & [h:m:s] & [$^\circ$:$\arcmin$:$\arcsec$] & [mag] & [mag] & [mag] & [mag] &
 [$\arcmin$] & \\
 & & & & & & & & \\[-1mm]
\hline
 & & & & & & & & \\[-1mm]
1 & 03:55:02.93 & $-$49:36:44.6 & 18.27 & 1.45 & 0.01 & 0.01 & 0.21 & RGB \\ 
2 & 03:55:06.36 & $-$49:36:09.5 & 19.44 & 1.07 & 0.01 & 0.01 & 1.01 & RGB \\ 
3 & 03:55:00.64 & $-$49:37:17.9 & 19.48 & 1.06 & 0.01 & 0.01 & 0.46 & RGB \\ 
4 & 03:55:02.04 & $-$49:36:51.7 & 19.51 & 1.02 & 0.01 & 0.01 & 0.07 & RGB \\ 
5 & 03:55:06.33 & $-$49:36:20.2 & 19.97 & 0.95 & 0.01 & 0.01 & 0.88 & RGB \\ 
6 & 03:55:01.82 & $-$49:36:39.9 & 20.12 & 0.92 & 0.01 & 0.01 & 0.27 & RGB \\ 
7 & 03:55:02.15 & $-$49:36:50.1 & 20.37 & 0.90 & 0.01 & 0.02 & 0.09 & RGB \\ 
8 & 03:55:01.79 & $-$49:41:06.0 & 20.43 & 0.73 & 0.01 & 0.01 & 4.18 & AGB \\ 
9 & 03:55:04.19 & $-$49:36:55.7 & 20.45 & 0.67 & 0.01 & 0.01 & 0.31 & AGB \\ 
10 & 03:55:06.78 & $-$49:37:02.1 & 20.52 & 0.75 & 0.01 & 0.02 & 0.74 & AGB \\ 
11 & 03:55:01.31 & $-$49:36:41.6 & 20.62 & 0.81 & 0.01 & 0.02 & 0.28 & RGB \\ 
12 & 03:55:05.70 & $-$49:37:25.9 & 20.63 & 0.67 & 0.01 & 0.01 & 0.75 & AGB \\ 
13 & 03:55:03.25 & $-$49:37:04.3 & 20.66 & 0.62 & 0.01 & 0.01 & 0.21 & HB \\ 
14 & 03:55:02.36 & $-$49:36:56.9 & 20.73 & 0.63 & 0.02 & 0.02 & 0.03 & HB \\ 
15 & 03:54:59.64 & $-$49:38:34.8 & 20.78 & 0.63 & 0.01 & 0.01 & 1.71 & HB \\ 
16 & 03:54:58.65 & $-$49:37:39.3 & 20.78 & 0.80 & 0.02 & 0.02 & 0.94 & RGB \\ 
17 & 03:55:02.50 & $-$49:37:00.7 & 20.80 & 0.65 & 0.03 & 0.04 & 0.09 & HB \\ 
18 & 03:54:59.63 & $-$49:36:45.8 & 20.82 & 0.80 & 0.01 & 0.02 & 0.46 & RGB \\ 
19 & 03:55:01.04 & $-$49:36:47.6 & 20.87 & 0.73 & 0.01 & 0.02 & 0.24 & RGB \\ 
20 & 03:54:59.02 & $-$49:36:44.9 & 20.87 & 0.79 & 0.01 & 0.02 & 0.56 & RGB \\ 
21 & 03:55:01.27 & $-$49:36:45.5 & 20.88 & 0.59 & 0.01 & 0.01 & 0.23 & HB \\ 
22 & 03:55:00.68 & $-$49:36:45.0 & 20.88 & 0.80 & 0.01 & 0.02 & 0.31 & RGB \\ 
23 & 03:55:04.07 & $-$49:37:17.0 & 20.88 & 0.64 & 0.02 & 0.02 & 0.46 & HB \\ 
24 & 03:55:02.75 & $-$49:38:43.6 & 20.89 & 0.59 & 0.01 & 0.01 & 1.80 & HB \\ 
25 & 03:55:01.93 & $-$49:36:54.6 & 20.89 & 0.52 & 0.02 & 0.02 & 0.06 & HB \\ 
26 & 03:54:59.91 & $-$49:38:01.7 & 20.89 & 0.82 & 0.01 & 0.02 & 1.17 & RGB \\ 
27 & 03:55:04.57 & $-$49:37:02.1 & 20.92 & 0.57 & 0.01 & 0.02 & 0.39 & HB \\ 
28 & 03:55:02.92 & $-$49:37:00.0 & 20.92 & 0.80 & 0.02 & 0.02 & 0.13 & RGB \\ 
29 & 03:54:59.82 & $-$49:36:56.1 & 20.92 & 0.54 & 0.01 & 0.02 & 0.40 & HB \\ 
30 & 03:55:03.96 & $-$49:36:51.9 & 20.93 & 0.54 & 0.01 & 0.02 & 0.28 & HB \\ 
31 & 03:55:03.63 & $-$49:36:45.1 & 20.94 & 0.56 & 0.02 & 0.02 & 0.28 & HB \\ 
32 & 03:55:02.85 & $-$49:36:49.7 & 20.94 & 0.64 & 0.02 & 0.02 & 0.13 & HB \\ 
33 & 03:55:02.65 & $-$49:36:44.0 & 20.94 & 0.53 & 0.02 & 0.02 & 0.20 & HB \\ 
34 & 03:54:57.27 & $-$49:36:08.0 & 20.96 & 0.57 & 0.01 & 0.01 & 1.13 & HB \\ 
35 & 03:54:59.62 & $-$49:36:40.7 & 20.96 & 0.54 & 0.01 & 0.01 & 0.50 & HB \\ 
36 & 03:55:01.82 & $-$49:37:14.8 & 20.98 & 0.55 & 0.02 & 0.02 & 0.33 & HB \\ 
37 & 03:54:43.93 & $-$49:34:34.2 & 20.98 & 0.75 & 0.02 & 0.02 & 3.80 & RGB \\ 
38 & 03:55:02.13 & $-$49:37:26.1 & 20.98 & 0.60 & 0.01 & 0.02 & 0.51 & HB \\ 
39 & 03:55:01.19 & $-$49:36:30.6 & 20.98 & 0.58 & 0.01 & 0.02 & 0.45 & HB \\ 
40 & 03:54:58.67 & $-$49:36:20.8 & 20.98 & 0.52 & 0.01 & 0.01 & 0.82 & HB \\ 
41 & 03:55:03.07 & $-$49:36:46.5 & 20.99 & 0.59 & 0.02 & 0.02 & 0.19 & HB \\ 
42 & 03:55:03.54 & $-$49:36:43.9 & 20.99 & 0.53 & 0.02 & 0.02 & 0.28 & HB \\ 
43 & 03:55:01.39 & $-$49:36:47.4 & 20.99 & 0.57 & 0.01 & 0.02 & 0.20 & HB \\ 
44 & 03:54:50.84 & $-$49:38:21.4 & 21.00 & 0.64 & 0.01 & 0.02 & 2.34 & HB \\ 
45 & 03:55:04.06 & $-$49:37:01.5 & 21.00 & 0.52 & 0.02 & 0.02 & 0.30 & HB \\ 
46 & 03:55:04.85 & $-$49:36:48.4 & 21.00 & 0.54 & 0.01 & 0.02 & 0.43 & HB \\ 
47 & 03:55:06.02 & $-$49:37:05.8 & 21.01 & 0.55 & 0.01 & 0.01 & 0.63 & HB \\ 
48 & 03:55:00.81 & $-$49:37:01.1 & 21.01 & 0.55 & 0.01 & 0.02 & 0.26 & HB \\ 
49 & 03:55:01.38 & $-$49:37:00.4 & 21.02 & 0.52 & 0.02 & 0.02 & 0.17 & HB \\ 
50 & 03:55:02.01 & $-$49:36:55.6 & 21.02 & 0.53 & 0.02 & 0.02 & 0.05 & HB \\ 
51 & 03:55:04.22 & $-$49:37:04.0 & 21.02 & 0.79 & 0.02 & 0.02 & 0.34 & RGB \\ 
52 & 03:55:00.59 & $-$49:36:50.9 & 21.02 & 0.55 & 0.02 & 0.02 & 0.29 & HB \\ 
53 & 03:54:59.20 & $-$49:36:10.3 & 21.02 & 0.53 & 0.02 & 0.01 & 0.90 & HB \\ 
54 & 03:55:01.43 & $-$49:36:35.8 & 21.02 & 0.52 & 0.02 & 0.02 & 0.36 & HB \\ 
55 & 03:55:02.35 & $-$49:36:55.9 & 21.03 & 0.72 & 0.03 & 0.03 & 0.01 & RGB \\ 
56 & 03:55:02.30 & $-$49:36:37.9 & 21.04 & 0.56 & 0.01 & 0.02 & 0.29 & HB \\ 
57 & 03:55:01.43 & $-$49:37:05.7 & 21.04 & 0.77 & 0.01 & 0.02 & 0.22 & RGB \\ 
58 & 03:55:07.14 & $-$49:36:39.7 & 21.04 & 0.76 & 0.01 & 0.02 & 0.83 & RGB \\ 
59 & 03:54:59.93 & $-$49:36:51.5 & 21.06 & 0.49 & 0.02 & 0.02 & 0.39 & HB \\ 
60 & 03:55:03.35 & $-$49:37:12.9 & 21.09 & 0.53 & 0.02 & 0.02 & 0.34 & HB \\ 
61 & 03:55:01.28 & $-$49:37:06.2 & 21.17 & 0.75 & 0.02 & 0.03 & 0.24 & RGB \\ 
62 & 03:54:58.25 & $-$49:38:14.7 & 21.17 & 0.71 & 0.01 & 0.02 & 1.47 & RGB \\ 
63 & 03:55:02.59 & $-$49:37:02.2 & 21.17 & 0.77 & 0.02 & 0.04 & 0.12 & RGB \\ 
64 & 03:55:03.77 & $-$49:38:03.4 & 21.30 & 0.41 & 0.02 & 0.02 & 1.16 & HB \\ 
65 & 03:55:02.97 & $-$49:37:13.1 & 21.31 & 0.68 & 0.03 & 0.02 & 0.31 & RGB \\ 
66 & 03:55:01.21 & $-$49:36:20.9 & 21.33 & 0.73 & 0.01 & 0.02 & 0.60 & RGB \\ 
67 & 03:55:02.67 & $-$49:36:46.7 & 21.37 & 0.69 & 0.02 & 0.03 & 0.16 & RGB \\ 
68 & 03:55:04.21 & $-$49:37:29.2 & 21.40 & 0.73 & 0.02 & 0.03 & 0.64 & RGB \\ 
69 & 03:55:03.66 & $-$49:36:48.4 & 21.41 & 0.79 & 0.02 & 0.03 & 0.25 & RGB \\ 
70 & 03:55:12.00 & $-$49:34:51.1 & 21.46 & 0.68 & 0.03 & 0.03 & 2.60 & RGB \\ 
71 & 03:55:01.91 & $-$49:36:27.5 & 21.47 & 0.73 & 0.02 & 0.03 & 0.47 & RGB \\ 
72 & 03:55:09.89 & $-$49:37:36.2 & 21.50 & 0.43 & 0.02 & 0.02 & 1.40 & HB \\ 
73 & 03:55:01.58 & $-$49:36:56.1 & 21.58 & 0.73 & 0.03 & 0.03 & 0.12 & RGB \\ 
74 & 03:55:03.45 & $-$49:36:04.8 & 21.67 & 0.76 & 0.02 & 0.02 & 0.87 & RGB \\ 
75 & 03:55:02.47 & $-$49:37:05.0 & 21.69 & 0.67 & 0.03 & 0.04 & 0.16 & RGB \\ 
76 & 03:55:03.23 & $-$49:36:51.6 & 21.73 & 0.70 & 0.04 & 0.04 & 0.16 & RGB \\ 
77 & 03:55:06.81 & $-$49:37:38.8 & 21.85 & 0.66 & 0.03 & 0.03 & 1.03 & RGB \\ 
78 & 03:55:02.10 & $-$49:37:21.2 & 21.85 & 0.75 & 0.03 & 0.04 & 0.43 & RGB \\ 
79 & 03:55:03.47 & $-$49:36:50.0 & 21.87 & 0.70 & 0.08 & 0.07 & 0.21 & RGB \\ 
80 & 03:55:01.81 & $-$49:37:03.9 & 21.88 & 0.78 & 0.04 & 0.04 & 0.16 & RGB \\ 
81 & 03:55:02.31 & $-$49:37:02.3 & 21.91 & 0.77 & 0.03 & 0.05 & 0.11 & RGB \\ 
82 & 03:54:50.34 & $-$49:34:53.5 & 21.95 & 0.72 & 0.03 & 0.04 & 2.81 & RGB \\ 
83 & 03:55:03.82 & $-$49:36:32.7 & 22.04 & 0.64 & 0.03 & 0.04 & 0.45 & RGB \\ 
84 & 03:55:02.54 & $-$49:37:19.0 & 22.06 & 0.68 & 0.03 & 0.04 & 0.39 & RGB \\ 
85 & 03:55:04.96 & $-$49:37:12.8 & 22.08 & 0.66 & 0.03 & 0.05 & 0.52 & RGB \\ 
86 & 03:55:02.63 & $-$49:37:18.5 & 22.14 & 0.71 & 0.04 & 0.04 & 0.39 & RGB \\ 
87 & 03:55:04.29 & $-$49:36:38.6 & 22.17 & 0.65 & 0.03 & 0.04 & 0.43 & RGB \\ 
88 & 03:55:00.41 & $-$49:37:00.4 & 22.22 & 0.66 & 0.03 & 0.04 & 0.32 & RGB \\ 
89 & 03:55:03.87 & $-$49:35:48.2 & 22.33 & 0.62 & 0.03 & 0.04 & 1.15 & RGB \\ 
90 & 03:54:59.31 & $-$49:36:15.6 & 22.34 & 0.65 & 0.04 & 0.05 & 0.82 & RGB \\ 
\hline
\end{tabular}
}
\end{table}

\begin{table}[h]
{\bf Table A.1. (continued)}\\[1mm]
{\tiny
\begin{tabular}{r@{\hspace{3mm}}c@{\hspace{3mm}}cc@{\hspace{3mm}}c@{\hspace{3mm}
}c@{\hspace{3mm}}ccc}
\hline
 & & & & & & & & \\[-1mm]
Id & $\alpha$(2000) & $\delta$(2000) & $V$ & $B-V$ & $\sigma_V$ & $\sigma_B$ &
 radius & Type \\
 & [h:m:s] & [$^\circ$:$\arcmin$:$\arcsec$] & [mag] & [mag] & [mag] & [mag] &
 [$\arcmin$] & \\
 & & & & & & & & \\[-1mm]
\hline
 & & & & & & & & \\[-1mm]
91 & 03:55:02.00 & $-$49:37:09.0 & 22.35 & 0.68 & 0.04 & 0.06 & 0.23 & RGB \\ 
92 & 03:55:04.18 & $-$49:37:46.5 & 22.36 & 0.64 & 0.04 & 0.05 & 0.90 & RGB \\ 
93 & 03:54:51.97 & $-$49:35:54.4 & 22.39 & 0.64 & 0.06 & 0.05 & 1.96 & RGB \\ 
94 & 03:54:58.27 & $-$49:36:28.0 & 22.45 & 0.62 & 0.04 & 0.04 & 0.80 & RGB \\ 
95 & 03:55:05.75 & $-$49:36:03.1 & 22.47 & 0.69 & 0.03 & 0.06 & 1.04 & RGB \\ 
96 & 03:55:02.60 & $-$49:37:11.6 & 22.53 & 0.68 & 0.04 & 0.06 & 0.27 & RGB \\ 
97 & 03:55:10.81 & $-$49:40:31.9 & 22.56 & 0.71 & 0.06 & 0.06 & 3.86 & RGB \\ 
98 & 03:55:02.88 & $-$49:36:31.9 & 22.58 & 0.65 & 0.04 & 0.05 & 0.40 & RGB \\ 
99 & 03:55:03.88 & $-$49:36:58.2 & 22.60 & 0.63 & 0.06 & 0.07 & 0.26 & RGB \\ 
100 & 03:54:59.98 & $-$49:36:53.9 & 22.61 & 0.68 & 0.05 & 0.06 & 0.38 & RGB \\ 
101 & 03:54:43.96 & $-$49:37:59.2 & 22.61 & 0.73 & 0.05 & 0.07 & 3.15 & RGB \\ 
102 & 03:54:59.13 & $-$49:37:12.9 & 22.67 & 0.63 & 0.06 & 0.07 & 0.59 & RGB \\ 
103 & 03:55:01.82 & $-$49:37:09.1 & 22.69 & 0.62 & 0.04 & 0.07 & 0.24 & RGB \\ 
104 & 03:55:22.06 & $-$49:36:08.5 & 22.69 & 0.63 & 0.05 & 0.07 & 3.30 & RGB \\ 
105 & 03:55:02.10 & $-$49:36:45.8 & 22.75 & 0.66 & 0.06 & 0.07 & 0.16 & RGB \\ 
106 & 03:54:49.76 & $-$49:37:22.8 & 22.77 & 0.59 & 0.07 & 0.06 & 2.08 & RGB \\ 
107 & 03:55:05.37 & $-$49:36:44.6 & 22.77 & 0.62 & 0.05 & 0.09 & 0.53 & RGB \\ 
108 & 03:55:02.05 & $-$49:37:01.3 & 22.78 & 0.60 & 0.05 & 0.10 & 0.10 & RGB \\ 
109 & 03:55:04.44 & $-$49:36:37.5 & 22.89 & 0.70 & 0.05 & 0.10 & 0.46 & RGB \\ 
110 & 03:55:01.53 & $-$49:37:20.8 & 22.91 & 0.67 & 0.06 & 0.08 & 0.44 & RGB \\ 
\hline
\end{tabular}
}
\end{table}

\begin{table}[h]
\caption{\label{tab6} List of probable member stars of Pal\,3, ordered with
increasing $V$ magnitude.}
{\tiny
\begin{tabular}{r@{\hspace{3mm}}c@{\hspace{3mm}}cc@{\hspace{3mm}}c@{\hspace{3mm}
}c@{\hspace{3mm}}ccc}
\hline
 & & & & & & & & \\[-1mm]
Id & $\alpha$(2000) & $\delta$(2000) & $V$ & $B-V$ & $\sigma_V$ & $\sigma_B$ &
 radius & Type \\
 & [h:m:s] & [$^\circ$:$\arcmin$:$\arcsec$] & [mag] & [mag] & [mag] & [mag] &
 [$\arcmin$] & \\
 & & & & & & & & \\[-1mm]
\hline
 & & & & & & & & \\[-1mm]
1 & 10:05:36.24 & 00:03:59.9 & 18.05 & 1.16 & 0.01 & 0.01 & 1.11 & RGB \\ 
2 & 10:05:31.57 & 00:04:17.0 & 18.25 & 1.13 & 0.02 & 0.01 & 0.10 & RGB \\ 
3 & 10:05:37.08 & 00:04:27.9 & 18.37 & 1.12 & 0.01 & 0.01 & 1.29 & RGB \\ 
4 & 10:05:29.71 & 00:05:39.2 & 18.64 & 0.95 & 0.02 & 0.01 & 1.46 & AGB? \\ 
5 & 10:05:31.05 & 00:04:17.3 & 18.76 & 1.00 & 0.01 & 0.01 & 0.23 & RGB \\ 
6 & 10:05:34.63 & 00:04:06.9 & 19.30 & 0.89 & 0.01 & 0.01 & 0.69 & RGB \\ 
7 & 10:05:21.69 & 00:07:35.2 & 19.30 & 0.81 & 0.02 & 0.01 & 4.17 & AGB? \\ 
8 & 10:05:31.84 & 00:04:16.5 & 19.34 & 0.86 & 0.02 & 0.01 & 0.04 & RGB \\ 
9 & 10:05:32.87 & 00:04:36.1 & 19.37 & 0.88 & 0.02 & 0.01 & 0.38 & RGB \\ 
10 & 10:05:33.19 & 00:03:57.1 & 19.42 & 0.88 & 0.02 & 0.01 & 0.47 & RGB \\ 
11 & 10:05:27.72 & 00:05:08.8 & 19.54 & 0.85 & 0.02 & 0.01 & 1.35 & RGB \\ 
12 & 10:05:31.12 & 00:04:16.9 & 19.59 & 0.89 & 0.02 & 0.01 & 0.21 & RGB \\ 
13 & 10:05:32.91 & 00:04:26.1 & 19.80 & 0.85 & 0.02 & 0.01 & 0.27 & RGB \\ 
14 & 10:05:32.59 & 00:04:32.3 & 19.85 & 0.82 & 0.02 & 0.01 & 0.29 & RGB \\ 
15 & 10:05:32.73 & 00:04:36.7 & 19.97 & 0.81 & 0.02 & 0.01 & 0.37 & RGB \\ 
16 & 10:05:29.92 & 00:04:53.9 & 20.04 & 0.77 & 0.04 & 0.03 & 0.78 & RGB \\ 
17 & 10:05:28.29 & 00:04:37.7 & 20.08 & 0.61 & 0.01 & 0.01 & 0.97 & RGB \\ 
18 & 10:05:23.96 & 00:04:24.5 & 20.12 & 0.71 & 0.01 & 0.01 & 2.00 & AGB \\ 
19 & 10:05:30.25 & 00:03:37.6 & 20.14 & 0.80 & 0.01 & 0.01 & 0.80 & RGB \\ 
20 & 10:05:30.36 & 00:05:29.6 & 20.18 & 0.59 & 0.02 & 0.01 & 1.26 & AGB \\ 
21 & 10:05:28.83 & 00:04:36.5 & 20.22 & 0.79 & 0.01 & 0.01 & 0.84 & RGB \\ 
22 & 10:05:28.79 & 00:03:45.5 & 20.22 & 0.78 & 0.01 & 0.01 & 0.96 & RGB \\ 
23 & 10:05:29.97 & 00:04:03.8 & 20.26 & 0.78 & 0.01 & 0.01 & 0.55 & RGB \\ 
24 & 10:05:30.44 & 00:03:38.1 & 20.27 & 0.28 & 0.01 & 0.01 & 0.77 & HB \\ 
25 & 10:05:34.35 & 00:04:39.2 & 20.27 & 0.65 & 0.01 & 0.01 & 0.70 & AGB \\ 
26 & 10:05:33.82 & 00:04:18.2 & 20.28 & 0.34 & 0.02 & 0.01 & 0.47 & HB \\ 
27 & 10:05:21.94 & 00:04:06.3 & 20.30 & 0.33 & 0.02 & 0.01 & 2.51 & HB \\ 
28 & 10:05:30.07 & 00:02:04.7 & 20.31 & 0.80 & 0.02 & 0.01 & 2.27 & RGB \\ 
29 & 10:05:31.49 & 00:04:27.7 & 20.31 & 0.76 & 0.02 & 0.01 & 0.20 & RGB \\ 
30 & 10:05:31.35 & 00:04:14.4 & 20.33 & 0.28 & 0.02 & 0.01 & 0.16 & HB \\ 
31 & 10:05:35.79 & 00:03:34.3 & 20.33 & 0.50 & 0.01 & 0.01 & 1.20 & HB \\ 
32 & 10:05:32.63 & 00:04:21.3 & 20.39 & 0.37 & 0.02 & 0.01 & 0.18 & HB \\ 
33 & 10:05:35.91 & 00:05:10.3 & 20.39 & 0.39 & 0.01 & 0.01 & 1.32 & HB \\ 
34 & 10:05:33.32 & 00:04:38.0 & 20.40 & 0.45 & 0.02 & 0.01 & 0.48 & HB \\ 
35 & 10:05:27.59 & 00:05:00.0 & 20.42 & 0.41 & 0.02 & 0.01 & 1.30 & HB \\ 
36 & 10:05:39.49 & 00:00:10.8 & 20.43 & 0.64 & 0.03 & 0.02 & 4.53 & AGB \\ 
37 & 10:05:34.47 & 00:03:57.1 & 20.43 & 0.45 & 0.01 & 0.01 & 0.72 & HB \\ 
38 & 10:05:27.77 & 00:04:35.5 & 20.43 & 0.43 & 0.01 & 0.01 & 1.09 & HB \\ 
39 & 10:05:30.44 & 00:04:22.9 & 20.44 & 0.46 & 0.02 & 0.01 & 0.39 & HB \\ 
40 & 10:05:28.24 & 00:03:34.3 & 20.45 & 0.29 & 0.01 & 0.01 & 1.18 & HB \\ 
41 & 10:05:29.61 & 00:04:21.3 & 20.45 & 0.51 & 0.01 & 0.01 & 0.59 & HB \\ 
42 & 10:05:34.00 & 00:04:27.4 & 20.45 & 0.49 & 0.01 & 0.01 & 0.54 & HB \\ 
43 & 10:05:35.85 & 00:04:15.8 & 20.46 & 0.44 & 0.01 & 0.01 & 0.98 & HB \\ 
44 & 10:05:29.65 & 00:04:12.0 & 20.47 & 0.35 & 0.04 & 0.01 & 0.58 & HB \\ 
45 & 10:05:27.39 & 00:05:09.7 & 20.47 & 0.38 & 0.02 & 0.01 & 1.43 & HB \\ 
46 & 10:05:31.53 & 00:04:40.5 & 20.47 & 0.45 & 0.01 & 0.01 & 0.39 & HB \\ 
47 & 10:05:28.80 & 00:04:10.9 & 20.48 & 0.51 & 0.01 & 0.01 & 0.80 & HB \\ 
48 & 10:05:29.12 & 00:03:45.8 & 20.49 & 0.49 & 0.02 & 0.01 & 0.89 & HB \\ 
49 & 10:05:35.95 & 00:05:05.6 & 20.50 & 0.48 & 0.01 & 0.01 & 1.28 & HB \\ 
50 & 10:05:31.82 & 00:04:25.4 & 20.51 & 0.75 & 0.02 & 0.01 & 0.13 & RGB \\ 
51 & 10:05:28.34 & 00:04:48.7 & 20.54 & 0.76 & 0.01 & 0.01 & 1.04 & RGB \\ 
52 & 10:05:36.69 & 00:03:15.7 & 20.54 & 0.39 & 0.01 & 0.01 & 1.58 & HB \\ 
53 & 10:05:31.54 & 00:04:21.7 & 20.54 & 0.41 & 0.02 & 0.01 & 0.12 & HB \\ 
54 & 10:05:38.09 & 00:04:53.3 & 20.61 & 0.74 & 0.01 & 0.01 & 1.64 & RGB \\ 
55 & 10:05:30.53 & 00:04:37.3 & 20.61 & 0.47 & 0.02 & 0.01 & 0.48 & HB \\ 
56 & 10:05:33.21 & 00:05:06.2 & 20.63 & 0.75 & 0.02 & 0.01 & 0.86 & RGB \\
57 & 10:05:35.28 & 00:03:43.1 & 20.70 & 0.40 & 0.01 & 0.01 & 1.02 & HB \\
58 & 10:05:21.84 & 00:01:05.8 & 20.70 & 0.49 & 0.04 & 0.03 & 4.08 & HB \\
59 & 10:05:33.41 & 00:04:28.5 & 20.70 & 0.42 & 0.02 & 0.01 & 0.40 & HB \\
60 & 10:05:32.37 & 00:04:12.1 & 20.72 & 0.72 & 0.02 & 0.01 & 0.14 & RGB \\
61 & 10:05:20.31 & 00:07:37.3 & 20.76 & 0.42 & 0.02 & 0.01 & 4.42 & HB \\
\hline
\end{tabular}
}
\end{table}

\begin{table}[h]
{\bf Table A.2. (continued)}\\[1mm]
{\tiny
\begin{tabular}{r@{\hspace{3mm}}c@{\hspace{3mm}}cc@{\hspace{3mm}}c@{\hspace{3mm}
}c@{\hspace{3mm}}ccc}
\hline
 & & & & & & & & \\[-1mm]
Id & $\alpha$(2000) & $\delta$(2000) & $V$ & $B-V$ & $\sigma_V$ & $\sigma_B$ &
 radius & Type \\
 & [h:m:s] & [$^\circ$:$\arcmin$:$\arcsec$] & [mag] & [mag] & [mag] & [mag] &
 [$\arcmin$] & \\
 & & & & & & & & \\[-1mm]
\hline
 & & & & & & & & \\[-1mm]
62 & 10:05:30.12 & 00:04:46.0 & 20.86 & 0.72 & 0.02 & 0.01 & 0.65 & RGB \\ 
63 & 10:05:30.87 & 00:04:39.9 & 20.92 & 0.69 & 0.02 & 0.02 & 0.45 & RGB \\ 
64 & 10:05:31.51 & 00:04:54.5 & 20.95 & 0.68 & 0.02 & 0.01 & 0.62 & RGB \\ 
65 & 10:05:31.88 & 00:06:17.2 & 20.96 & 0.66 & 0.02 & 0.02 & 1.99 & RGB \\ 
66 & 10:05:25.51 & 00:04:15.7 & 21.01 & 0.71 & 0.02 & 0.01 & 1.61 & RGB \\ 
67 & 10:05:35.85 & 00:04:54.5 & 21.02 & 0.70 & 0.01 & 0.01 & 1.15 & RGB \\ 
68 & 10:05:32.48 & 00:04:13.6 & 21.03 & 0.72 & 0.02 & 0.02 & 0.15 & RGB \\ 
69 & 10:05:42.34 & 00:01:04.2 & 21.05 & 0.69 & 0.04 & 0.03 & 4.14 & RGB \\ 
70 & 10:05:33.10 & 00:04:11.9 & 21.07 & 0.70 & 0.02 & 0.02 & 0.31 & RGB \\ 
71 & 10:05:29.63 & 00:04:23.8 & 21.10 & 0.71 & 0.02 & 0.02 & 0.59 & RGB \\ 
72 & 10:05:33.12 & 00:04:36.1 & 21.14 & 0.73 & 0.03 & 0.02 & 0.42 & RGB \\ 
73 & 10:05:30.90 & 00:04:48.5 & 21.17 & 0.66 & 0.04 & 0.03 & 0.57 & RGB \\ 
74 & 10:05:33.18 & 00:03:59.9 & 21.22 & 0.63 & 0.02 & 0.02 & 0.43 & RGB \\ 
75 & 10:05:30.90 & 00:04:33.9 & 21.28 & 0.63 & 0.02 & 0.02 & 0.37 & RGB \\ 
76 & 10:05:31.21 & 00:05:03.3 & 21.32 & 0.70 & 0.02 & 0.02 & 0.78 & RGB \\ 
77 & 10:05:32.30 & 00:03:55.3 & 21.36 & 0.73 & 0.02 & 0.03 & 0.39 & RGB \\ 
78 & 10:05:37.93 & 00:04:07.8 & 21.42 & 0.72 & 0.02 & 0.01 & 1.50 & RGB \\ 
79 & 10:05:32.77 & 00:04:15.2 & 21.43 & 0.66 & 0.02 & 0.02 & 0.21 & RGB \\ 
80 & 10:05:29.94 & 00:04:28.5 & 21.48 & 0.68 & 0.02 & 0.01 & 0.53 & RGB \\ 
81 & 10:05:33.44 & 00:04:25.2 & 21.51 & 0.70 & 0.02 & 0.02 & 0.39 & RGB \\ 
82 & 10:05:25.75 & 00:04:20.8 & 21.52 & 0.66 & 0.02 & 0.01 & 1.55 & RGB \\ 
83 & 10:05:30.78 & 00:04:29.1 & 21.57 & 0.61 & 0.03 & 0.02 & 0.35 & RGB \\ 
84 & 10:05:30.89 & 00:03:51.0 & 21.58 & 0.65 & 0.03 & 0.02 & 0.52 & RGB \\ 
85 & 10:05:38.12 & 00:05:00.2 & 21.63 & 0.67 & 0.02 & 0.01 & 1.70 & RGB \\ 
86 & 10:05:29.29 & 00:04:24.8 & 21.66 & 0.69 & 0.02 & 0.02 & 0.68 & RGB \\ 
87 & 10:05:29.48 & 00:03:20.0 & 21.70 & 0.68 & 0.02 & 0.02 & 1.15 & RGB \\ 
88 & 10:05:32.08 & 00:04:16.7 & 21.71 & 0.65 & 0.03 & 0.02 & 0.04 & RGB \\ 
89 & 10:05:32.02 & 00:04:51.4 & 21.73 & 0.66 & 0.02 & 0.02 & 0.56 & RGB \\ 
90 & 10:05:28.88 & 00:04:12.1 & 21.76 & 0.66 & 0.03 & 0.02 & 0.77 & RGB \\ 
91 & 10:05:28.96 & 00:04:20.4 & 21.77 & 0.69 & 0.02 & 0.01 & 0.75 & RGB \\ 
92 & 10:05:31.12 & 00:04:32.2 & 21.78 & 0.62 & 0.03 & 0.03 & 0.31 & RGB \\ 
93 & 10:05:32.00 & 00:04:22.0 & 21.80 & 0.69 & 0.03 & 0.03 & 0.07 & RGB \\ 
94 & 10:05:30.37 & 00:04:03.1 & 21.81 & 0.65 & 0.05 & 0.04 & 0.47 & RGB \\ 
95 & 10:05:33.34 & 00:04:18.2 & 21.83 & 0.65 & 0.04 & 0.02 & 0.35 & RGB \\ 
96 & 10:05:31.61 & 00:04:12.2 & 21.88 & 0.62 & 0.04 & 0.03 & 0.13 & RGB \\ 
97 & 10:05:32.43 & 00:03:59.8 & 21.89 & 0.59 & 0.03 & 0.02 & 0.33 & RGB \\ 
98 & 10:05:29.36 & 00:08:33.4 & 21.92 & 0.66 & 0.04 & 0.03 & 4.31 & RGB \\ 
99 & 10:05:31.72 & 00:04:19.8 & 21.92 & 0.67 & 0.03 & 0.02 & 0.07 & RGB \\ 
100 & 10:05:35.42 & 00:03:07.8 & 21.92 & 0.62 & 0.02 & 0.02 & 1.46 & RGB \\ 
101 & 10:05:33.85 & 00:06:17.2 & 21.93 & 0.64 & 0.03 & 0.03 & 2.04 & RGB \\ 
102 & 10:05:33.59 & 00:02:53.0 & 21.93 & 0.65 & 0.02 & 0.01 & 1.48 & RGB \\ 
103 & 10:05:30.03 & 00:04:28.1 & 21.94 & 0.70 & 0.02 & 0.02 & 0.51 & RGB \\ 
104 & 10:05:32.34 & 00:04:21.4 & 21.98 & 0.63 & 0.03 & 0.03 & 0.11 & RGB \\ 
105 & 10:05:33.53 & 00:04:20.9 & 21.99 & 0.68 & 0.03 & 0.02 & 0.40 & RGB \\ 
106 & 10:05:29.06 & 00:04:29.4 & 22.02 & 0.60 & 0.03 & 0.02 & 0.75 & RGB \\ 
107 & 10:05:37.11 & 00:04:58.6 & 22.02 & 0.65 & 0.02 & 0.02 & 1.46 & RGB \\ 
108 & 10:05:36.87 & 00:03:42.5 & 22.04 & 0.64 & 0.03 & 0.03 & 1.36 & RGB \\ 
109 & 10:05:33.07 & 00:04:09.4 & 22.04 & 0.69 & 0.03 & 0.02 & 0.32 & RGB \\ 
110 & 10:05:23.81 & 00:04:22.7 & 22.07 & 0.64 & 0.02 & 0.02 & 2.04 & RGB \\ 
111 & 10:05:40.33 & 00:09:14.9 & 22.09 & 0.60 & 0.04 & 0.03 & 5.37 & RGB \\ 
112 & 10:05:27.96 & 00:04:26.0 & 22.10 & 0.71 & 0.03 & 0.02 & 1.01 & RGB \\ 
113 & 10:05:21.85 & 00:07:41.0 & 22.10 & 0.58 & 0.04 & 0.03 & 4.22 & RGB \\ 
114 & 10:05:41.53 & 00:05:45.4 & 22.14 & 0.67 & 0.05 & 0.03 & 2.80 & RGB \\ 
115 & 10:05:33.00 & 00:06:34.6 & 22.16 & 0.59 & 0.04 & 0.03 & 2.29 & RGB \\ 
116 & 10:05:31.59 & 00:03:28.9 & 22.17 & 0.59 & 0.03 & 0.02 & 0.82 & RGB \\ 
117 & 10:05:29.98 & 00:04:27.1 & 22.18 & 0.64 & 0.04 & 0.04 & 0.52 & RGB \\ 
118 & 10:05:28.18 & 00:04:35.3 & 22.18 & 0.57 & 0.03 & 0.02 & 0.98 & RGB \\ 
119 & 10:05:28.65 & 00:05:48.8 & 22.18 & 0.65 & 0.03 & 0.02 & 1.72 & RGB \\ 
120 & 10:05:34.50 & 00:04:42.8 & 22.20 & 0.70 & 0.03 & 0.02 & 0.76 & RGB \\ 
121 & 10:05:30.95 & 00:04:12.4 & 22.21 & 0.59 & 0.04 & 0.03 & 0.27 & RGB \\ 
122 & 10:05:33.34 & 00:05:35.4 & 22.23 & 0.70 & 0.04 & 0.04 & 1.34 & RGB \\ 
123 & 10:05:32.15 & 00:04:51.1 & 22.23 & 0.63 & 0.04 & 0.02 & 0.55 & RGB \\ 
124 & 10:05:30.30 & 00:04:07.6 & 22.23 & 0.63 & 0.04 & 0.03 & 0.45 & RGB \\ 
125 & 10:05:33.96 & 00:03:30.5 & 22.24 & 0.63 & 0.04 & 0.03 & 0.94 & RGB \\ 
126 & 10:05:27.88 & 00:08:29.6 & 22.26 & 0.60 & 0.05 & 0.04 & 4.31 & RGB \\ 
127 & 10:05:31.40 & 00:03:25.4 & 22.26 & 0.64 & 0.05 & 0.03 & 0.89 & RGB \\ 
128 & 10:05:31.84 & 00:04:42.6 & 22.27 & 0.70 & 0.04 & 0.03 & 0.41 & RGB \\ 
129 & 10:05:30.64 & 00:03:16.8 & 22.30 & 0.63 & 0.04 & 0.03 & 1.07 & RGB \\ 
130 & 10:05:31.42 & 00:03:36.1 & 22.30 & 0.65 & 0.04 & 0.03 & 0.71 & RGB \\ 
131 & 10:05:32.78 & 00:04:09.7 & 22.30 & 0.64 & 0.06 & 0.04 & 0.25 & RGB \\ 
132 & 10:05:29.59 & 00:04:26.3 & 22.34 & 0.64 & 0.03 & 0.03 & 0.61 & RGB \\ 
133 & 10:05:31.92 & 00:05:08.9 & 22.35 & 0.69 & 0.05 & 0.04 & 0.85 & RGB \\ 
134 & 10:05:34.04 & 00:03:44.3 & 22.35 & 0.56 & 0.04 & 0.04 & 0.77 & RGB \\ 
135 & 10:05:28.01 & 00:04:06.1 & 22.36 & 0.66 & 0.04 & 0.03 & 1.00 & RGB \\ 
136 & 10:05:32.55 & 00:03:34.7 & 22.39 & 0.67 & 0.04 & 0.03 & 0.74 & RGB \\ 
137 & 10:05:32.12 & 00:04:57.7 & 22.40 & 0.71 & 0.06 & 0.05 & 0.66 & RGB \\ 
138 & 10:05:29.26 & 00:03:44.5 & 22.41 & 0.62 & 0.04 & 0.03 & 0.88 & RGB \\ 
139 & 10:05:30.79 & 00:04:24.1 & 22.44 & 0.62 & 0.05 & 0.03 & 0.31 & RGB \\ 
140 & 10:05:36.89 & 00:06:35.1 & 22.44 & 0.56 & 0.05 & 0.03 & 2.60 & RGB \\ 
141 & 10:05:37.92 & 00:02:41.4 & 22.45 & 0.57 & 0.07 & 0.07 & 2.20 & RGB \\ 
142 & 10:05:30.71 & 00:03:46.7 & 22.48 & 0.66 & 0.04 & 0.04 & 0.61 & RGB \\ 
143 & 10:05:32.34 & 00:04:05.6 & 22.48 & 0.62 & 0.04 & 0.04 & 0.23 & RGB \\ 
144 & 10:05:33.16 & 00:04:33.6 & 22.48 & 0.71 & 0.05 & 0.04 & 0.40 & RGB \\ 
145 & 10:05:33.91 & 00:04:48.9 & 22.48 & 0.59 & 0.03 & 0.02 & 0.71 & RGB \\ 
146 & 10:05:30.26 & 00:03:45.6 & 22.52 & 0.69 & 0.05 & 0.04 & 0.69 & RGB \\ 
147 & 10:05:37.41 & 00:06:11.1 & 22.52 & 0.56 & 0.05 & 0.03 & 2.33 & RGB \\ 
148 & 10:05:28.19 & 00:03:26.2 & 22.52 & 0.66 & 0.04 & 0.04 & 1.28 & RGB \\ 
149 & 10:05:24.98 & 00:04:14.3 & 22.52 & 0.69 & 0.04 & 0.04 & 1.74 & RGB \\ 
150 & 10:05:32.97 & 00:04:37.8 & 22.53 & 0.55 & 0.05 & 0.03 & 0.42 & RGB \\ 
151 & 10:05:32.30 & 00:03:22.6 & 22.60 & 0.67 & 0.05 & 0.03 & 0.93 & RGB \\ 
152 & 10:05:27.74 & 00:03:19.7 & 22.62 & 0.59 & 0.04 & 0.04 & 1.43 & RGB \\ 
153 & 10:05:29.79 & 00:03:46.0 & 22.62 & 0.67 & 0.05 & 0.04 & 0.76 & RGB \\ 
\hline
\end{tabular}
}
\end{table}

\begin{table}[h]
\caption{\label{tab7} List of probable member stars of Pal\,14, ordered with
increasing $V$ magnitude.}
{\tiny
\begin{tabular}{r@{\hspace{3mm}}c@{\hspace{3mm}}cc@{\hspace{3mm}}c@{\hspace{3mm}
}c@{\hspace{3mm}}ccc}
\hline
 & & & & & & & & \\[-1mm]
Id & $\alpha$(2000) & $\delta$(2000) & $V$ & $B-V$ & $\sigma_V$ & $\sigma_B$ &
 radius & Type \\
 & [h:m:s] & [$^\circ$:$\arcmin$:$\arcsec$] & [mag] & [mag] & [mag] & [mag] &
 [$\arcmin$] & \\
 & & & & & & & & \\[-1mm]
\hline
 & & & & & & & & \\[-1mm]
1 & 16:11:05.81 & 14:57:45.1 & 17.37 & 1.39 & 0.02 & 0.01 & 1.28 & RGB \\ 
2 & 16:10:58.73 & 14:56:48.7 & 17.77 & 1.29 & 0.02 & 0.01 & 0.80 & RGB \\ 
3 & 16:10:54.90 & 14:58:36.7 & 18.23 & 1.07 & 0.01 & 0.01 & 1.80 & AGB? \\ 
4 & 16:11:04.98 & 14:53:32.3 & 18.48 & 1.10 & 0.02 & 0.01 & 4.07 & RGB \\ 
5 & 16:10:59.24 & 14:57:22.5 & 18.52 & 0.99 & 0.01 & 0.01 & 0.35 & AGB? \\ 
6 & 16:10:53.36 & 14:56:45.4 & 18.56 & 1.04 & 0.01 & 0.01 & 1.90 & RGB \\ 
7 & 16:10:54.04 & 14:57:05.6 & 18.70 & 1.03 & 0.01 & 0.01 & 1.64 & RGB \\ 
8 & 16:10:56.90 & 14:57:56.5 & 18.84 & 0.99 & 0.01 & 0.01 & 1.02 & RGB \\ 
9 & 16:11:01.40 & 14:57:60.0 & 19.05 & 0.99 & 0.01 & 0.01 & 0.56 & RGB \\ 
10 & 16:11:05.89 & 14:58:43.2 & 19.19 & 0.93 & 0.02 & 0.01 & 1.78 & RGB \\ 
11 & 16:11:06.00 & 14:55:39.1 & 19.37 & 0.95 & 0.01 & 0.01 & 2.23 & RGB \\ 
12 & 16:10:56.21 & 14:56:32.7 & 19.41 & 0.90 & 0.01 & 0.01 & 1.42 & RGB \\ 
13 & 16:10:56.98 & 14:56:25.8 & 19.44 & 0.92 & 0.01 & 0.01 & 1.37 & RGB \\ 
14 & 16:10:59.24 & 14:57:19.7 & 19.50 & 0.90 & 0.02 & 0.01 & 0.37 & RGB \\ 
15 & 16:10:55.84 & 14:57:43.4 & 19.60 & 0.69 & 0.01 & 0.01 & 1.19 & AGB \\ 
16 & 16:10:59.62 & 15:01:32.9 & 19.68 & 0.81 & 0.02 & 0.01 & 4.09 & RGB \\ 
17 & 16:11:00.58 & 14:56:59.1 & 19.76 & 0.86 & 0.01 & 0.01 & 0.48 & RGB \\ 
18 & 16:11:03.91 & 14:56:34.3 & 19.77 & 0.59 & 0.01 & 0.01 & 1.19 & AGB \\ 
19 & 16:11:01.01 & 14:55:23.1 & 19.78 & 0.69 & 0.01 & 0.01 & 2.09 & AGB \\ 
20 & 16:11:07.42 & 14:57:43.2 & 19.86 & 0.86 & 0.01 & 0.01 & 1.65 & RGB \\ 
21 & 16:10:52.81 & 14:55:38.9 & 19.87 & 0.84 & 0.01 & 0.01 & 2.62 & RGB \\ 
22 & 16:10:56.17 & 14:55:48.1 & 19.88 & 0.46 & 0.01 & 0.01 & 1.99 & HB \\ 
23 & 16:11:02.52 & 14:56:49.7 & 19.88 & 0.62 & 0.01 & 0.01 & 0.78 & HB \\ 
24 & 16:11:10.09 & 14:55:54.9 & 19.89 & 0.88 & 0.02 & 0.01 & 2.76 & RGB \\ 
25 & 16:11:04.45 & 14:56:14.4 & 19.90 & 0.65 & 0.01 & 0.01 & 1.53 & HB \\ 
26 & 16:10:59.32 & 14:57:25.9 & 19.92 & 0.84 & 0.02 & 0.01 & 0.32 & RGB \\ 
27 & 16:11:16.22 & 15:00:07.1 & 19.92 & 0.47 & 0.02 & 0.01 & 4.60 & HB \\ 
28 & 16:10:59.89 & 14:53:27.8 & 19.92 & 0.92 & 0.03 & 0.02 & 4.01 & RGB \\ 
29 & 16:11:12.60 & 14:56:55.2 & 19.93 & 0.46 & 0.02 & 0.01 & 2.94 & HB \\ 
30 & 16:11:13.56 & 14:57:58.2 & 19.95 & 0.85 & 0.02 & 0.01 & 3.16 & RGB \\ 
31 & 16:11:01.64 & 14:59:47.8 & 19.95 & 0.80 & 0.02 & 0.02 & 2.34 & RGB \\ 
32 & 16:10:56.39 & 14:57:25.1 & 19.96 & 0.84 & 0.01 & 0.02 & 1.03 & RGB \\ 
33 & 16:11:09.37 & 14:57:37.5 & 19.97 & 0.61 & 0.02 & 0.01 & 2.11 & HB \\ 
34 & 16:10:58.62 & 14:57:40.3 & 19.97 & 0.83 & 0.01 & 0.01 & 0.53 & RGB \\ 
35 & 16:11:09.57 & 15:01:07.7 & 19.98 & 0.58 & 0.02 & 0.02 & 4.25 & HB \\ 
36 & 16:10:59.15 & 14:57:38.0 & 19.98 & 0.76 & 0.01 & 0.01 & 0.40 & HB? \\ 
37 & 16:10:56.20 & 14:57:34.4 & 19.99 & 0.61 & 0.01 & 0.01 & 1.08 & HB \\ 
38 & 16:11:01.76 & 14:58:06.8 & 19.99 & 0.61 & 0.01 & 0.01 & 0.70 & HB \\ 
39 & 16:10:59.72 & 14:57:45.6 & 20.00 & 0.58 & 0.01 & 0.01 & 0.37 & HB \\ 
40 & 16:10:59.15 & 14:55:01.8 & 20.03 & 0.67 & 0.02 & 0.02 & 2.46 & HB \\ 
41 & 16:11:00.81 & 14:57:53.4 & 20.03 & 0.58 & 0.02 & 0.01 & 0.42 & HB \\ 
42 & 16:10:53.24 & 14:57:50.2 & 20.03 & 0.55 & 0.01 & 0.01 & 1.83 & HB \\ 
43 & 16:10:53.08 & 14:56:22.9 & 20.03 & 0.55 & 0.02 & 0.01 & 2.13 & HB \\ 
44 & 16:10:55.09 & 14:57:44.3 & 20.04 & 0.60 & 0.01 & 0.01 & 1.37 & HB \\ 
45 & 16:11:01.49 & 14:57:46.0 & 20.04 & 0.58 & 0.02 & 0.01 & 0.36 & HB \\ 
46 & 16:11:08.53 & 14:53:09.0 & 20.04 & 0.50 & 0.02 & 0.01 & 4.72 & HB \\ 
47 & 16:10:53.97 & 14:56:35.4 & 20.06 & 0.55 & 0.01 & 0.02 & 1.84 & HB \\ 
48 & 16:11:03.24 & 14:57:44.9 & 20.06 & 0.62 & 0.01 & 0.01 & 0.69 & HB \\ 
49 & 16:11:16.01 & 14:54:36.2 & 20.07 & 0.64 & 0.03 & 0.02 & 4.69 & HB \\ 
50 & 16:11:02.83 & 14:57:09.0 & 20.07 & 0.56 & 0.01 & 0.01 & 0.62 & HB \\ 
51 & 16:10:59.37 & 14:57:43.5 & 20.07 & 0.57 & 0.01 & 0.01 & 0.40 & HB \\ 
52 & 16:10:53.27 & 14:56:27.2 & 20.07 & 0.54 & 0.01 & 0.01 & 2.05 & HB \\ 
53 & 16:11:00.38 & 14:57:40.7 & 20.08 & 0.55 & 0.02 & 0.01 & 0.22 & HB \\ 
54 & 16:11:09.51 & 14:59:43.9 & 20.08 & 0.57 & 0.02 & 0.01 & 3.12 & HB \\ 
55 & 16:10:59.25 & 14:57:14.0 & 20.08 & 0.58 & 0.01 & 0.01 & 0.41 & HB \\ 
56 & 16:10:59.65 & 14:56:43.4 & 20.11 & 0.59 & 0.01 & 0.02 & 0.78 & HB \\ 
57 & 16:11:00.94 & 14:58:50.3 & 20.11 & 0.41 & 0.01 & 0.01 & 1.37 & HB \\ 
58 & 16:11:01.91 & 14:57:53.2 & 20.11 & 0.84 & 0.01 & 0.02 & 0.52 & RGB \\ 
59 & 16:10:56.97 & 14:57:28.0 & 20.11 & 0.56 & 0.01 & 0.01 & 0.89 & HB \\ 
60 & 16:10:57.39 & 14:56:01.3 & 20.13 & 0.86 & 0.02 & 0.02 & 1.65 & RGB \\ 
61 & 16:11:01.71 & 14:57:36.9 & 20.14 & 0.57 & 0.01 & 0.01 & 0.30 & HB \\ 
62 & 16:10:59.47 & 14:56:03.6 & 20.19 & 0.67 & 0.01 & 0.01 & 1.44 & HB \\ 
63 & 16:11:13.75 & 14:54:18.3 & 20.24 & 0.88 & 0.04 & 0.02 & 4.48 & RGB \\ 
64 & 16:10:59.47 & 14:58:02.1 & 20.28 & 0.84 & 0.02 & 0.02 & 0.63 & RGB \\ 
65 & 16:10:59.65 & 14:55:42.4 & 20.30 & 0.79 & 0.01 & 0.02 & 1.78 & RGB \\ 
66 & 16:10:58.97 & 14:55:28.4 & 20.33 & 0.82 & 0.02 & 0.01 & 2.04 & RGB \\ 
67 & 16:10:54.49 & 14:57:19.5 & 20.34 & 0.82 & 0.01 & 0.01 & 1.49 & RGB \\ 
68 & 16:10:59.66 & 14:58:18.5 & 20.38 & 0.83 & 0.02 & 0.02 & 0.87 & RGB \\ 
69 & 16:11:01.04 & 14:58:32.6 & 20.39 & 0.83 & 0.02 & 0.02 & 1.08 & RGB \\ 
70 & 16:11:15.80 & 14:57:55.6 & 20.41 & 0.91 & 0.02 & 0.02 & 3.69 & RGB \\ 
71 & 16:10:57.31 & 14:59:21.7 & 20.42 & 0.44 & 0.02 & 0.01 & 2.06 & HB?\\ 
72 & 16:11:08.15 & 14:57:40.5 & 20.48 & 0.77 & 0.02 & 0.02 & 1.82 & RGB \\ 
73 & 16:11:07.45 & 14:58:30.7 & 20.51 & 0.79 & 0.02 & 0.02 & 1.95 & RGB \\ 
74 & 16:11:03.28 & 14:58:04.1 & 20.54 & 0.81 & 0.02 & 0.02 & 0.87 & RGB \\ 
75 & 16:10:56.94 & 14:57:10.5 & 20.58 & 0.78 & 0.02 & 0.02 & 0.94 & RGB \\ 
76 & 16:10:57.67 & 14:57:19.1 & 20.60 & 0.74 & 0.01 & 0.02 & 0.73 & RGB \\ 
77 & 16:10:59.59 & 14:57:20.1 & 20.67 & 0.74 & 0.01 & 0.02 & 0.29 & RGB \\ 
78 & 16:11:11.43 & 14:58:09.6 & 20.67 & 0.81 & 0.02 & 0.02 & 2.69 & RGB \\ 
79 & 16:11:00.93 & 14:57:31.7 & 20.83 & 0.74 & 0.02 & 0.03 & 0.09 & RGB \\ 
80 & 16:10:58.97 & 14:53:23.0 & 20.85 & 0.81 & 0.03 & 0.02 & 4.11 & RGB \\ 
81 & 16:11:07.10 & 14:58:17.8 & 20.85 & 0.78 & 0.02 & 0.02 & 1.76 & RGB \\ 
82 & 16:11:01.63 & 14:55:27.6 & 20.90 & 0.77 & 0.02 & 0.03 & 2.02 & RGB \\ 
83 & 16:11:00.90 & 14:58:06.8 & 20.97 & 0.70 & 0.02 & 0.03 & 0.65 & RGB \\ 
84 & 16:10:59.36 & 14:56:58.3 & 20.98 & 0.76 & 0.02 & 0.03 & 0.59 & RGB \\ 
85 & 16:11:17.29 & 14:56:19.9 & 21.00 & 0.72 & 0.02 & 0.02 & 4.18 & RGB \\ 
86 & 16:10:54.40 & 14:58:31.1 & 21.02 & 0.74 & 0.02 & 0.02 & 1.84 & RGB \\ 
87 & 16:10:54.08 & 14:58:33.6 & 21.05 & 0.77 & 0.02 & 0.03 & 1.93 & RGB \\ 
88 & 16:11:02.26 & 14:58:25.2 & 21.10 & 0.82 & 0.02 & 0.04 & 1.03 & RGB \\ 
89 & 16:11:06.41 & 14:53:22.1 & 21.12 & 0.85 & 0.02 & 0.04 & 4.33 & RGB \\ 
90 & 16:11:09.75 & 14:57:55.8 & 21.18 & 0.75 & 0.02 & 0.04 & 2.25 & RGB \\ 
\hline
\end{tabular}
}
\end{table}

\begin{table}[h]
{\bf Table A.3. (continued)}\\[1mm]
{\tiny
\begin{tabular}{r@{\hspace{3mm}}c@{\hspace{3mm}}cc@{\hspace{3mm}}c@{\hspace{3mm}
}c@{\hspace{3mm}}ccc}
\hline
 & & & & & & & & \\[-1mm]
Id & $\alpha$(2000) & $\delta$(2000) & $V$ & $B-V$ & $\sigma_V$ & $\sigma_B$ &
 radius & Type \\
 & [h:m:s] & [$^\circ$:$\arcmin$:$\arcsec$] & [mag] & [mag] & [mag] & [mag] &
 [$\arcmin$] & \\
 & & & & & & & & \\[-1mm]
\hline
 & & & & & & & & \\[-1mm]
91 & 16:11:05.62 & 14:58:45.8 & 21.18 & 0.69 & 0.02 & 0.03 & 1.77 & RGB \\ 
92 & 16:11:07.25 & 14:57:11.9 & 21.21 & 0.79 & 0.02 & 0.04 & 1.62 & RGB \\ 
93 & 16:11:02.05 & 14:58:21.3 & 21.21 & 0.70 & 0.02 & 0.04 & 0.95 & RGB \\ 
94 & 16:10:57.26 & 14:57:35.4 & 21.23 & 0.71 & 0.02 & 0.03 & 0.83 & RGB \\ 
95 & 16:11:00.23 & 14:58:02.6 & 21.23 & 0.72 & 0.02 & 0.03 & 0.58 & RGB \\ 
96 & 16:10:59.37 & 14:57:35.3 & 21.24 & 0.72 & 0.02 & 0.03 & 0.33 & RGB \\ 
97 & 16:11:03.92 & 14:56:21.7 & 21.29 & 0.70 & 0.03 & 0.03 & 1.36 & RGB \\ 
98 & 16:10:59.12 & 14:56:52.0 & 21.30 & 0.74 & 0.02 & 0.04 & 0.70 & RGB \\ 
99 & 16:11:05.63 & 14:59:20.3 & 21.30 & 0.76 & 0.03 & 0.05 & 2.22 & RGB \\ 
100 & 16:11:00.25 & 14:58:12.4 & 21.35 & 0.78 & 0.03 & 0.04 & 0.75 & RGB \\ 
101 & 16:11:08.20 & 14:58:53.5 & 21.40 & 0.69 & 0.02 & 0.03 & 2.31 & RGB \\ 
102 & 16:10:59.70 & 14:57:29.8 & 21.41 & 0.77 & 0.02 & 0.04 & 0.23 & RGB \\ 
103 & 16:11:01.68 & 14:57:09.4 & 21.42 & 0.73 & 0.02 & 0.04 & 0.40 & RGB \\ 
104 & 16:10:58.82 & 14:57:17.6 & 21.43 & 0.76 & 0.02 & 0.04 & 0.47 & RGB \\ 
105 & 16:11:00.37 & 14:57:45.6 & 21.47 & 0.73 & 0.03 & 0.04 & 0.30 & RGB \\ 
106 & 16:11:01.58 & 14:58:49.8 & 21.48 & 0.74 & 0.02 & 0.04 & 1.38 & RGB \\ 
107 & 16:10:57.76 & 14:55:27.3 & 21.54 & 0.81 & 0.03 & 0.05 & 2.13 & RGB \\ 
108 & 16:11:00.59 & 14:57:03.1 & 21.56 & 0.67 & 0.03 & 0.05 & 0.42 & RGB \\ 
109 & 16:11:01.72 & 14:57:56.4 & 21.59 & 0.75 & 0.02 & 0.04 & 0.54 & RGB \\ 
110 & 16:11:03.46 & 14:56:35.5 & 21.65 & 0.75 & 0.02 & 0.05 & 1.11 & RGB \\ 
111 & 16:11:02.28 & 14:58:13.4 & 21.66 & 0.63 & 0.03 & 0.04 & 0.85 & RGB \\ 
112 & 16:11:00.03 & 14:57:27.7 & 21.72 & 0.78 & 0.03 & 0.05 & 0.15 & RGB \\ 
113 & 16:10:55.84 & 14:56:54.1 & 21.72 & 0.73 & 0.03 & 0.06 & 1.29 & RGB \\ 
114 & 16:11:15.66 & 14:59:26.6 & 21.74 & 0.74 & 0.03 & 0.06 & 4.13 & RGB \\ 
115 & 16:11:02.13 & 14:57:17.1 & 21.75 & 0.79 & 0.05 & 0.05 & 0.40 & RGB \\ 
116 & 16:11:01.79 & 14:55:48.3 & 21.76 & 0.68 & 0.03 & 0.05 & 1.68 & RGB \\ 
117 & 16:11:00.66 & 14:57:03.8 & 21.78 & 0.68 & 0.03 & 0.04 & 0.40 & RGB \\ 
118 & 16:10:58.93 & 14:57:53.1 & 21.80 & 0.70 & 0.02 & 0.05 & 0.59 & RGB \\ 
119 & 16:11:01.50 & 14:58:28.3 & 21.82 & 0.71 & 0.03 & 0.05 & 1.02 & RGB \\ 
120 & 16:10:57.79 & 14:56:23.9 & 21.83 & 0.69 & 0.03 & 0.05 & 1.27 & RGB \\ 
121 & 16:11:04.03 & 15:02:01.1 & 21.84 & 0.78 & 0.03 & 0.06 & 4.62 & RGB \\ 
122 & 16:10:58.53 & 14:57:02.2 & 21.85 & 0.65 & 0.03 & 0.05 & 0.67 & RGB \\ 
123 & 16:11:01.37 & 14:57:40.2 & 21.87 & 0.78 & 0.03 & 0.05 & 0.27 & RGB \\ 
124 & 16:10:55.65 & 14:57:12.9 & 21.92 & 0.64 & 0.03 & 0.06 & 1.23 & RGB \\ 
125 & 16:11:01.20 & 14:58:37.3 & 21.92 & 0.66 & 0.02 & 0.04 & 1.16 & RGB \\ 
126 & 16:11:00.14 & 14:58:22.3 & 21.93 & 0.67 & 0.04 & 0.06 & 0.91 & RGB \\ 
127 & 16:11:05.39 & 14:56:59.9 & 21.94 & 0.69 & 0.04 & 0.07 & 1.24 & RGB \\ 
128 & 16:10:55.79 & 14:58:02.7 & 22.00 & 0.72 & 0.04 & 0.06 & 1.31 & RGB \\ 
129 & 16:10:59.39 & 15:01:17.9 & 22.01 & 0.62 & 0.04 & 0.05 & 3.84 & RGB \\ 
130 & 16:11:00.63 & 14:58:27.8 & 22.01 & 0.78 & 0.03 & 0.07 & 0.99 & RGB \\ 
131 & 16:10:59.19 & 14:56:02.5 & 22.02 & 0.71 & 0.03 & 0.08 & 1.47 & RGB \\ 
132 & 16:11:10.01 & 14:58:51.7 & 22.06 & 0.75 & 0.02 & 0.04 & 2.66 & RGB \\ 
133 & 16:11:02.13 & 14:57:03.0 & 22.08 & 0.72 & 0.03 & 0.07 & 0.55 & RGB \\ 
134 & 16:10:58.95 & 14:58:04.7 & 22.10 & 0.61 & 0.04 & 0.07 & 0.74 & RGB \\ 
135 & 16:10:55.42 & 14:57:20.6 & 22.11 & 0.69 & 0.04 & 0.07 & 1.27 & RGB \\ 
136 & 16:10:57.33 & 14:57:23.4 & 22.14 & 0.61 & 0.05 & 0.07 & 0.81 & RGB \\ 
137 & 16:11:03.62 & 14:57:22.5 & 22.15 & 0.76 & 0.04 & 0.07 & 0.72 & RGB \\ 
138 & 16:11:00.66 & 14:56:51.7 & 22.18 & 0.68 & 0.04 & 0.08 & 0.61 & RGB \\ 
139 & 16:11:00.83 & 14:58:22.2 & 22.24 & 0.69 & 0.04 & 0.08 & 0.90 & RGB \\ 
140 & 16:11:01.28 & 14:56:51.2 & 22.27 & 0.71 & 0.04 & 0.11 & 0.63 & RGB \\ 
141 & 16:10:54.85 & 15:00:45.2 & 22.29 & 0.70 & 0.05 & 0.09 & 3.57 & RGB \\ 
142 & 16:11:03.71 & 14:57:16.6 & 22.30 & 0.78 & 0.04 & 0.09 & 0.76 & RGB \\ 
143 & 16:11:04.49 & 14:57:38.2 & 22.31 & 0.70 & 0.05 & 0.09 & 0.94 & RGB \\ 
\hline
\end{tabular}
}
\end{table}

\enddocument